\documentclass[twocolumn,trackchanges]{aastex701}
\UseRawInputEncoding
\usepackage{booktabs,units}    
\usepackage{tabularx}   
\usepackage{makecell}
\usepackage{orcidlink}

\usepackage{xcolor}
\usepackage{physics}
\usepackage[thinc]{esdiff}
\usepackage{graphicx,subcaption} 
\usepackage{multirow}
\definecolor{Red}{named}{red}

\newcommand{\SPA}{School of Physics and Astronomy, Monash University, Clayton VIC 3800, Australia}
\newcommand{\OzGravMonash}{OzGrav: The ARC Centre of Excellence for Gravitational Wave Discovery, Clayton VIC 3800, Australia}
\newcommand{\OzGravMelbs}{OzGrav: The ARC Centre of Excellence for Gravitational Wave Discovery, Melbourne VIC 3010, Australia}
\newcommand{\SoPUM}{School of Physics, The University of Melbourne, VIC 3010, Australia}

\begin{document}

\title{The properties of the first continuous gravitational waves: \\
Optimizing pulsar timing observations for supermassive black hole binary detection}

\author{Blaze Houlden}
\affiliation{\SoPUM}
\affiliation{\OzGravMelbs}
\email{bhoulden@student.unimelb.edu.au}
\orcidlink{0009-0008-9886-0282}

\author{Eric Thrane}
\affiliation{\SPA}
\affiliation{\OzGravMonash}
\email{}

\author{Katie Auchettl}
\affiliation{\SoPUM}
\affiliation{\OzGravMelbs}
\email{}

\begin{abstract}
Several pulsar timing arrays have reported evidence for a nanohertz gravitational-wave background.
This background is thought to arise from the superposition of signals from a population of inspiraling supermassive black hole binaries.
A key test of this interpretation is the search for individually resolvable binaries whose signals emerge above this background.
Here we investigate this possibility by addressing two related questions: (1) what are the most probable properties of the first resolvable binary; and (2) how can pulsar timing observations be optimized to maximize its prospects for electromagnetic identification?
By simulating populations of supermassive black hole binaries constrained by evidence for the gravitational-wave background and studying the loudest binary in each simulation, we infer the expected properties of the first individually resolvable systems.
If MeerKAT continues its current observing strategy for a total of 9 years, the probability of detecting an individual binary is $\approx10-27\%$. 
Improving the timing precision by a factor of two changes this probability to $\approx11-28\%$, whereas increasing the cadence by a factor of four gives $\approx12-30\%$.
Combining both improvements yields a detection probability of $\approx12-33\%$.
Furthermore, implementing a high-cadence observing campaign preferentially resolves higher-frequency systems ($\gtrsim\unit[10]{nHz}$), whose shorter orbital periods make them considerably easier to identify through electromagnetic observations than lower-frequency binaries that repeat infrequently.
We discuss prospects of electromagnetic follow-up of gravitational-wave resolved binaries with current high-cadence all-sky optical surveys.

\end{abstract}


\section{Introduction} 
\label{sec:intro}

Supermassive black holes, with mass $\gtrsim 10^6\;\rm{M}_{\odot}$, are observed to reside in the center of nearly every galaxy \citep{KormendyHo2013}.
Over cosmic time, galaxy mergers are an inevitable consequence of hierarchical structure formation, naturally leading to the formation of supermassive black hole binaries (SMBHBs; \citealt{begelman_massive_1980}).

Supermassive black hole binary candidates have been electromagnetically observed at parsec- to kiloparsec-scale separations as ``dual active galactic nuclei" (e.g., \citealt{chen_varstrometry_2022}); however, there are no confirmed SMBHB systems with sub-parsec separations.
These close separations make it difficult to electromagnetically resolve SMBHB systems at distances of megaparsecs \citep[see][and references therein]{dorazio_observational_2023}.
However, Earth-sized very-long baseline interferometers, such as the Event Horizon Telescope, have sufficient angular resolution to resolve SMBHB systems with parsec-scale separation at distances of $\lesssim \unit[100]{Mpc}$---provided they are sufficiently radio bright.

Another approach to identifying SMBHB candidates is to search for periodic or quasi-periodic variability in the light curves of active galactic nuclei that may arise from binary orbital motion. 
This method has unearthed several candidates, the best known of which is OJ 287, a quasar with quasi-periodic outbursts at approximately 12-year intervals (\citealt{sillanpaa_oj_1988, valtonen_massive_2008, valtonen_primary_2016, valtonen_identifying_2025, gupta_multiband_2026}).
In the binary interpretation, this periodicity has been attributed to the secondary black hole passing through the accretion disk of the primary (e.g., \citealt{valtonen_primary_2016, dey_unique_2019, ressler_black_2025, garain_massive_2026, chitan_long-term_2026}).
The true nature of quasi-periodic variability candidates, however, is uncertain, because the mechanisms by which binary systems may produce periodic variability are not yet well understood.
Moreover, stochastic variability from the accretion disk around a single supermassive black hole may also masquerade as quasi-periodicity on timescales shorter or comparable to a binary's proposed orbital time (e.g., \citealt{vaughan_false_2016, liu_supermassive_2019, witt_quasars_2022, el-badry_active_2025, molina_search_2025}).
Additionally, alternative processes, such as cold spots in an individual supermassive black hole accretion disk, may also lead to quasi-periodic signals (\citealt{kovacevic_optical_2019}).

Another notable candidate, hosted in the elliptical galaxy 3C~66B, was thought to harbor an SMBHB system from the detection of periodic variability in the radio core of the galaxy (\citealt{sudou_orbital_2003, jenet_constraining_2004, iguchi_very_2010}). 
However, the inferred binary parameters imply a stochastic background that is inconsistent with current PTA results (\citealt{zhu_minimum_2019}).
More recently, pulsar timing array observations have placed upper limits on the binary mass that are significantly lower than those inferred from the electromagnetic observations (\citealt{tremblay_multi-messenger_2025}).
These examples illustrate the difficulty in detecting an SMBHB candidate using electromagnetic observations alone.
The electromagnetic signatures expected from these objects remain uncertain, while several astrophysical processes unrelated to binary evolution can produce variability that mimics the anticipated signatures of SMBHB systems. 
Consequently, current electromagnetic candidates remain difficult to interpret unambiguously.

Gravitational-wave observations provide an independent means of detecting and characterizing sub-parsec SMBHB systems. 
Pulsar timing arrays (PTAs) are sensitive to gravitational waves in the nanohertz frequency band, making them uniquely suited to probe the inspiral of the most massive binary systems at distances farther than $\sim100$~Mpc. 
As gravitational waves propagate through the Earth-pulsar system, they induce perturbations in the pulse arrival times. 
The perturbations are correlated between pairs of millisecond pulsars depending on their angular separation. 
For an isotropic stochastic gravitational-wave background, these timing residuals exhibit the characteristic Hellings--Downs angular correlation (\citealt{hellings_upper_1983}).

The Australian Parkes PTA (PPTA; \citealt{reardon_search_2023}), the European and Indian PTA (EPTA and  InPTA; \citealt{antoniadis_second_2023}), the North American Nanohertz Observatory for Gravitational waves (NANOGrav; \citealt{agazie_nanograv_2023}), the Chinese PTA (CPTA; \citealt{xu_searching_2023}), and the MeerKAT PTA (MPTA; \citealt{miles_meerkat_2025}) independently reported statistically significant ($2-4\sigma$) evidence for a  Hellings--Downs correlated signal, suggesting a stochastic gravitational-wave background in the nanohertz frequency band. 
The leading astrophysical interpretation is that this signal arises from the superposition of gravitational waves emitted by a cosmological population of inspiralling supermassive black hole binaries (\citealt{rajagopal_ultra--low-frequency_1995}), although alternative cosmological origins --- including primordial gravitational waves following inflation (\citealt{grishchuk_primordial_1976, starobinsky_new_1980, linde_new_1982}), cosmological phase transitions (\citealt{caprini_detection_2010, caprini_cosmological_2018, caprini_detecting_2020}), and cosmic strings (\citealt{vilenkin_gravitational_1981, vilenkin_cosmic_1985, vilenkin_cosmic_1994, hindmarsh_cosmic_1995}) --- remain possible.

The strength of the gravitational-wave background is often parameterized with a dimensionless \textit{characteristic strain} amplitude, $A$, which is defined such that the power spectral density ($P(f)$) of the pulsar timing residuals\footnote{The units of the power spectral density $P(f)$ are $\unit[]{ns^2/Hz}$.} is given by
\begin{align} \label{eq:PSD}
    P(f) = & \frac{A^2}{12 \pi^2} f^{-11/3} (\unit[1]{yr})^{4/3} .
\end{align}
Here, $f$ is the frequency, and we use a reference frequency of $\unit[1]{yr^{-1}}$ to remain consistent with the PTA literature.

Current measurements of $A$ from PTAs range from $(2.1-4.1)\times 10^{-15}$ (\citealt{reardon_search_2023, antoniadis_second_2023, agazie_nanograv_2023, miles_meerkat_2025}).
These measurements are consistent with some theoretical predictions for the stochastic background arising from a population of SMBHB systems, although there are significant theoretical uncertainties in the predicted amplitude.
Theoretical predictions for $A$ span orders of magnitude, from early predictions of ${\cal O}(10^{-16})$ (\citealt{rajagopal_ultra--low-frequency_1995,dvorkin_nightmare_2017}), to more recent, but differing estimates of $10^{-15} - 2 \times 10^{-14}$ (\citealt{mcwilliams_gravitational_2014}; see Table A1 in \citealt{agazie_nanograv_2023-2} for a thorough overview of theoretical estimates).
The broad contrast between the various theoretical estimates and the current PTA results highlights the substantial uncertainties in modeling the SMBHB population and the stochastic background.
This motivates the need for further work to bridge the experimental and theoretical divide.

Pulsar timing arrays may be able to resolve the loudest individual SMBHB systems contributing to the stochastic background.
In the nanohertz frequency range, SMBHB systems are expected to evolve slowly, completing a small number of orbital cycles over several years. 
As a result, one expects these sources to exhibit negligible frequency evolution during their early inspiral phase. 
Their gravitational-wave emission can therefore be approximated as quasi-periodic, and they are commonly referred to as \textit{continuous-wave} sources.
To date, no PTA collaborations have reported statistically significant ($>5\sigma$) evidence of an individual continuous-wave source, and instead searches have placed upper limits on the amplitudes of these signals  (\citealt{agazie_nanograv_2023-1, antoniadis_second_2023, zhao_searching_2025, tian_targeted_2025, agarwal_nanograv_2026}).

As PTA sensitivity improves, a key objective is detecting individual continuous-wave sources.
The identification of a continuous-wave source consistent with an SMBHB would provide direct evidence that at least some part of the stochastic background originates from SMBHB systems.
Resolving individual SMBHB systems in PTA data could also enable the independent confirmation of electromagnetic candidate SMBHB systems found through optical variability and spectroscopic signatures. 
Beyond helping to determine the astrophysical origin of the stochastic gravitational wave background, the detection of individual SMBHB systems will allow us to probe the properties of these merging binaries, such as the demographics and evolution of SMBHB systems, revealing information about the galaxy merger process through cosmic time.

This information cannot be extracted from the stochastic background alone. 
For example, the background may be composed of many lighter SMBHB systems or relatively fewer heavier SMBHB systems that can produce similar amplitudes, leading to degeneracies between these cosmological models. 
Resolving individual SMBHB systems would help break these degeneracies and thereby constrain the astrophysical processes governing the SMBHB population.

This work has two goals: (1) to characterize the properties of SMBHB systems detectable with the MeerKAT PTA, and (2) to investigate how pulsar timing observations can be optimized to maximize the probability of detecting SMBHB systems as early as possible.
In Section \ref{sec:method} we describe the methods used to achieve these goals.
In Section \ref{sec:resolved}, we discuss the most likely properties of the first resolvable SMBHB. 
In Section \ref{sec:PTA_opt}, we investigate how different observing strategies affect the likelihood of detection of continuous-wave sources.
In Section \ref{sec:NG_comp}, we examine the SMBHB candidates found in \citet{agarwal_nanograv_2026}.
In Section \ref{sec:disc}, we discuss the implications of our work on current and future continuous-wave searches.
In Section \ref{sec:conc}, we state the key conclusions from our work.

\section{Method}  \label{sec:method}
\subsection{Signal Model}\label{sec:pta_signal}
In pulsar timing, pulsars are detectors that respond to the influence of a metric perturbation, the gravitational-wave strain, which is modeled in the $I^{\rm{th}}$ pulsar as
\begin{equation} \label{eq:resid}
    \delta t_I = M\epsilon + n_{\rm{white}} + n_{\rm{red}} + s.
\end{equation}
Here $M$ is the design matrix containing the timing model information, $\epsilon$ is a vector of offsets from the timing model, $n_{\rm{white}}$ describes the pulsar white noise processes, $n_{\mathrm{red}}$ the intrinsic pulsar red noise, and $s$ is the gravitational-wave signal.
White noise has the same power across all frequencies, whereas red noise has more power at low frequencies and can occur due to rotational instabilities in pulsars, denoted $n_{\mathrm{red}}$, or from a stochastic background, $s$.

The signal from an individual binary, denoted by $i$, can be expressed as 
\begin{equation} \label{eq:ind_sig}
    \begin{split}
        s_i(t) =&\; F^{+}(\hat{\Omega}) \left[ s_{+,i}(t_p) - s_{+,i}(t) \right] \\  &+ F^{\times}(\hat{\Omega})\left[ s_{\times,i}(t_p) - s_{\times,i}(t) \right],
    \end{split}
\end{equation}
where $t$ is the time measured at the solar system barycenter, and $t_p$ is the time at the pulsar, requiring knowledge of the sky location and distance to the pulsar, $s_{+/\times,\,i}(t)$ is the Earth-term signal and $s_{+/\times,\,i}(t_p)$ is the pulsar-term signal,  and $F^{+/ \times}(\hat{\Omega})$ are the antenna pattern functions that depend on the propagation direction of the gravitational wave, $\hat{\Omega}$, and the gravitational-wave polarization.
General relativity allows two polarizations of gravitational waves, known as the ``plus" (+) and ``cross" ($\times$) modes.

For the $I^{\rm{th}}$ pulsar, described with position vector, $\hat{p}_I$, the antenna pattern functions are defined as
\begin{equation} \label{eq:ant_pat}
    F_{I}^{A}(\hat{\Omega}) = \frac{1}{2}\frac{\hat{p}_{I}^{\mu} \hat{p}_{I}^{\nu}}{1 + \hat{\Omega}\cdot \hat{p}_{I}} e^{A}_{\mu\nu}(\hat{\Omega}),
\end{equation} 
where $e^{A}_{\mu\nu}$ are the polarization tensors of the gravitational wave for the $A = +, \times$ polarization states:
\begin{equation}
    e^{+}_{\mu\nu}(\hat{\Omega}) = \hat{m}_{\mu}\hat{m}_{\nu} - \hat{n}_{\mu}\hat{n}_{\nu},
\end{equation}
\begin{equation}
    e^{\times}_{\mu\nu}(\hat{\Omega}) = \hat{m}_{\mu}\hat{n}_{\nu} + \hat{n}_{\mu}\hat{m}_{\nu}.
\end{equation}
Both $\hat{m}$, $\hat{n}$ are unit vectors orthogonal to each other and $\hat{\Omega}$, are defined as
\begin{align}
    \hat{\Omega} & =-(\sin\theta\cos\phi)\hat{x} -(\sin\theta\sin\phi)\hat{y} - (\cos\theta)\hat{z}, \label{eq:omega}\\
    \hat{m}      & =(\sin\phi)\hat{x}-(\cos\phi)\hat{y},                                      \label{eq:m}    \\
    \hat{n}      & =-(\cos\theta\cos\phi)\hat{x} - (\cos\theta\sin\phi)\hat{y} + (\sin\theta)\hat{z}. \label{eq:n}
\end{align}
Here we use the basis vectors of the solar system barycenter coordinate system, $\hat{x}$, $\hat{y}$, and $\hat{z}$.
Similar to \citet{romano_answers_2024}, we use a minus sign in Eq.~\ref{eq:omega} to represent that $\theta$ and $\phi$ are the sky coordinates of the gravitational-wave source.\footnote{
A word of caution: in some works, $\theta$ and $\phi$ are defined as the sky coordinates of the propagation direction, which is reversed from the \citet{romano_answers_2024} convention we use here.}
The unit vector toward the pulsar 
\begin{equation}
    \hat{p}_I = (\sin\theta_I\cos\phi_I)\hat{x} +(\sin\theta_I\sin\phi_I)\hat{y} + (\cos\theta_I)\hat{z}, 
\end{equation}
is defined using the sky coordinates for the $I^{\mathrm{th}}$ pulsar, $(\theta_I, \phi_I)$.

Although the pulsar terms improve constraints on resolving the sky location, distance, and chirp mass of SMBHB systems \citep{corbin_pulsar_2010, lee_gravitational_2011}, the measured distances to the pulsars in the array are insufficiently precise for our method.
We thus neglect them in this work, introducing more noise into the calculation.
Therefore, Equation \ref{eq:ind_sig} becomes
\begin{equation} \label{eq:signal_i}
    \begin{split}
        s_i(t) \approx\; F^{+}(\hat{\Omega})  s_{+,i}(t)  +  F^{\times}(\hat{\Omega}) s_{\times,i}(t).
    \end{split}
\end{equation}

The Earth terms with $\times$ and $+$ polarizations for a circular-orbit, non-evolving binary can be expressed as
\begin{equation}
    s_{+,i}(t) = \frac{h_0}{2\pi f}(1 + \cos^2\iota)\sin(2\pi f t + \phi_0),
\end{equation}
\begin{equation}
    s_{\times,i}(t) = \frac{h_0}{2\pi f}2\cos\iota\cos(2\pi f t + \phi_0),
\end{equation}
where $f$ is the gravitational-wave frequency of the binary, $\iota$ is the angle of inclination of the binary, $\phi_0$ is some initial orbital phase, and $h_0$ is the dimensionless strain amplitude
\begin{equation} \label{eq:h0}
    \begin{split}
        h_0 =&\; 2\frac{(G\mathcal{M})^{5/3}}{c^4 D_{\rm{comov}}}(2\pi f_{{\mathrm{orb}}, r})^{2/3} \\ \equiv&\; 2\frac{(G\mathcal{M}_{z})^{5/3}}{c^4D_{L}} (2\pi f_{{\mathrm{orb}}})^{2/3},
    \end{split}
\end{equation}
where $\mathcal{M} = (m_1m_2)^{3/2}/(m_1 + m_2)^{1/5}$ is the chirp mass with $m_1$, $m_2$ the masses of the primary and secondary black holes, $\mathcal{M}_z = \mathcal{M}(1 + z)$ is the redshifted chirp mass, due to the cosmological redshift of the binary, $z$, $f_{{\mathrm{orb}}, r} = f_{{\mathrm{orb}}} / (1 + z)$ is the rest-frame orbital frequency, $D_{\rm{comov}}$ is the binary's comoving distance and $D_{L} = (1 + z)D_{\rm{comov}}$ is the binary's luminosity distance.
The gravitational-wave frequency of an SMBHB system in a circular orbit is simply twice the orbital frequency:  $f = 2 f_{\mathrm{orb}}$.

The total signal in the $I^{\mathrm{th}}$ pulsar, from our population of binaries, is the sum of contributions from each binary: 
\begin{equation}
    s(t) = \sum^{N_{\mathrm{binaries}}}_{i} s_i(t).
\end{equation}
This describes the superposition of all individual SMBHB gravitational-wave sources that form a stochastic background.

\subsection{Population Synthesis} \label{sec:pop_synth}

Each SMBHB system is characterized by: chirp mass, $\mathcal{M}$, mass ratio, $q=m_2/m_1 \leq 1$, comoving distance, $D_{\mathrm{comov}}$, observed gravitational-wave frequency, $f$, angle of inclination, $\iota$, initial orbital phase, $\phi_0$, and sky location (right ascension and declination in astronomical terms).
To synthesize a single SMBHB system, we sample from distributions for each of these components. 
\citet{agazie_nanograv_2023-2} use \texttt{holodeck} for their population synthesis, and parametrize their fiducial model for the primary mass following a power law that is exponentially suppressed above a characteristic mass, $m_{\rm{char}}$.
We follow the same parametrization:
\begin{equation} \label{eq:mass_prob}
    p(m_1) \propto 
    \begin{cases}
        m_1^{-\alpha} \exp(-m_1/m_{\rm{char}}),  & m_1 \in (m_{\rm{min}}, m_{\rm{max}}) \\
        0, & \text{elsewhere},
    \end{cases}
\end{equation}
for a black hole mass range of $(m_{\rm{min}}, m_{\rm{max}}) = (10^{7.5}\,\mathrm{M}_{\odot}, 10^{12.5}\,\mathrm{M}_{\odot}$).
We use $\alpha = -1.21 - 0.03z$ determined empirically from observations \citep[][]{chen_constraining_2019}, following the fiducial galaxy-merger number-density model of \citet{agazie_nanograv_2023-2}.
We adopt this parametrization and vary $m_{\rm char}$ to consider different cosmological population models.

The distribution of SMBHB mass ratio $q = m_2/m_1$ is poorly understood.
Following \citet{sesana_testing_2018}, we sample from a uniform distribution over $(q_{\mathrm{min}}, q_{\mathrm{max}}) = (0.05, 1.0)$.

The source-frame gravitational-wave frequency, $f_{\rm{s}}$, follows $p(f_{\rm{s}}) \propto f_s^{-11/3}$, as expected from a population of circular, gravitational-wave-driven inspiralling SMBHB systems (e.g., \citealt{smith_optimal_2018}). 
The observer-frame gravitational-wave frequency is given by $f = f_{\rm{s}}/(1 + z)$.
We parametrize our probability distribution by shifting to the observer frame using:
\begin{equation} \label{eq:f_prob}
    p(f|z) \propto \begin{cases}
        f^{-11/3}(1 + z)^{-8/3} & f \in (f_{\rm{min}}, f_{\rm{max}}) \;\mathrm{Hz} \\
        0, & \rm{elsewhere}
    \end{cases}\;\;.
\end{equation}
We adopt a frequency range of $(f_{\rm min}, f_{\rm max}) = (2\,\mathrm{nHz}, 300\,\mathrm{nHz})$, chosen to overlap with the MPTA observing band. 
High-redshift binaries are strongly suppressed, as they are redshifted below the PTA band.

We sample distance from a distribution that is uniform in comoving volume to a redshift of $z=2$.
We assume that SMBHB systems are isotropically distributed on the sky, and assign each binary a random initial orbital phase. 

We generate our population of SMBHB systems by repeatedly sampling from these distributions, forming populations that contain $4.5\times10^4-2\times10^8$ binaries.

We consider three scenarios, with parameters listed in Table \ref{tab:scenarios}:
\begin{itemize}
    \item \textbf{Light.} The stochastic background is due to a large number of relatively light binaries.
    \item \textbf{Intermediate.} The stochastic background is due to an intermediate number of binaries with an intermediate mass distribution.
    \item \textbf{Heavy.} The stochastic background is due to a relatively small number of relatively massive binaries.

\end{itemize}
Each of these models can produce identical results for the stochastic background signal, but the continuous-wave signal strength will vary due to the mass dependence in Equations \ref{eq:signal_i} and \ref{eq:h0}.

\begin{table} 
\caption{Parameter configurations for the three cosmological models of supermassive black hole binary populations considered in this work. }
    \centering
    \begin{tabular}{c|c|c}
   Scenario   & $m_{\mathrm{char}}$ [$\mathrm{M}_{\odot}$] & $N_{\mathrm{binaries}}$\\ 
   \hline 
   Light & $10^{8.7}$ & $6.0\times 10^{6} - 2.0\times10^{8}$\\
   Intermediate  & $10^{9.0}$ & $2.5\times 10^{6} - 1.0\times10^{8}$\\
   Heavy & $10^{9.3}$ & $4.5\times 10^{4} - 2.3\times10^{6}$\\
   
\end{tabular}
\label{tab:scenarios}
\end{table}

\subsubsection{PTA Optimal Statistic}
\label{subsec:pta_opt_stat}

We use the frequentist \textit{optimal statistic} to assess the significance of a stochastic gravitational-wave background in the pulsar data (\citealt{anholm_optimal_2009, demorest_limits_2013, chamberlin_time-domain_2015, vigeland_noise-marginalized_2018}).
The optimal statistic describes the cross-correlations in the pulsar data through the pulsar pair correlation coefficients
\begin{equation}
    \rho_{IJ} = \frac{\boldsymbol{\delta t_I}^T \mathbf{P}_I^{-1} \Tilde{\boldsymbol{\Phi}}_{IJ} \mathbf{P}_{J}^{-1} \boldsymbol{\delta t_J} }{\rm{Tr}\; \mathbf{P}_I^{-1} \Tilde{\boldsymbol{\Phi}}_{IJ} \mathbf{P}_{J}^{-1} \Tilde{\boldsymbol{\Phi}}_{JI}  },
\end{equation}
where $\boldsymbol{\delta t_I}^T$ is the vector of all residuals of the $I^{\rm{th}}$ pulsar from Eq.~\ref{eq:resid} and $\mathbf{P}_I = \langle \boldsymbol{\delta t_I} \boldsymbol{\delta t_I}^T \rangle$ is their total auto-covariance matrix.
Meanwhile, $\Tilde{\boldsymbol{\Phi}}_{IJ}$ is the cross-variance matrix of the pulsar pair, defined as 
\begin{align}
A^2\Gamma(\xi_{IJ})\Tilde{\boldsymbol{\Phi}}_{IJ} = \boldsymbol{\Phi}_{IJ} = \langle r_{I}r_{J}^{T}\rangle ,    
\end{align}
where $\xi_{IJ}$ is the angular separation of the $I^{\rm{th}}$ and $J^{\rm{th}}$ pulsars and  $\Gamma(\xi_{IJ})$ is the Hellings-Downs correlation coefficient (\citealt{hellings_upper_1983})
\begin{equation}
    \Gamma(\xi_{IJ}) = \frac{3}{2} \left\{ \frac{1}{3} + x\left[\ln\left(x\right) - \frac{1}{6}\right] \right\} + \frac{1}{2}\delta_{IJ},
\end{equation}
where $\delta_{IJ}$ normalizes this to unity when pulsar $I$ is pulsar $J$, and $x = (1 - \cos\xi_{IJ})/2$.
This angular correlation is used to differentiate the stochastic background from a spatially-uncorrelated red noise common process.
The associated variance for $\rho_{IJ}$ is defined as 
\begin{equation}
    \sigma_{IJ}^2 = (\rm{Tr}\; \mathbf{P}_I^{-1} \Tilde{\boldsymbol{\Phi}}_{IJ} \mathbf{P}_{J}^{-1} \Tilde{\boldsymbol{\Phi}}_{JI} )^{-1}.
\end{equation}

Similar to \citet{agazie_nanograv_2023-1}, we define the optimal statistic signal-to-noise ratio, $(\mathrm{S/N})_{\mathrm{SGWB}}$, as
\begin{equation}
    (\mathrm{S/N})_{\mathrm{SGWB}} = \frac{\sum_{I>J} \rho_{IJ}\Gamma(\xi_{IJ})/\sigma_{IJ}^2}{\left[\sum_{I>J}\Gamma^2(\xi_{aIJ})/\sigma_{IJ}^2 \right]^{1/2}}.
\end{equation}

\subsubsection{Synthesizing an MPTA-consistent SMBHB population} \label{sec:MPTA_consis}
We construct our simulation to mimic features of the MPTA 4.5-year data release \citep{miles_meerkat_2024}.
For the 4.5-year MPTA analysis, \citet{miles_meerkat_2025} modeled the common-spectrum (stochastic background) noise processes using the first 120 Fourier-frequency bins in each pulsar. 
We adopt the same choice for consistency; however, we find that it does not significantly affect our results.
Using \texttt{libstempo}, we set the original pulsar timing residuals to zero and inject realizations of white, red, chromatic, and dispersion-measure noise using the pulsar-specific noise parameters reported by \citet{miles_meerkat_2025}. 
We exclude PSR J1804--2858 because its large noise parameters prevented reliable recovery of physical values. 
We then inject the signal from each SMBHB system in our population into each pulsar following Equation~\ref{eq:signal_i}.

To synthesize an SMBHB population consistent with the $3.0\,\sigma-3.4\,\sigma$ evidence for Hellings-Downs correlations presented by \citet{miles_meerkat_2025}, we use the optimal statistic implementation in \texttt{enterprise\_extensions} to evaluate the $(\mathrm{S/N})_{\mathrm{SGWB}}$ for our simulated population.
We add binaries until we obtain a signal-to-noise ratio $3.0 \leq (\mathrm{S/N})_{\mathrm{SGWB}} \leq 3.4$ consistent with the results from \citet{miles_meerkat_2025}.
To gather a statistical ensemble for each scenario, we simulated 500 populations, each consistent with the stochastic background.

Finally, we simulate a further 4.5 years of observations for each simulation, with SMBHB signals injected, to produce a synthetic 9.0-year MPTA data set.
We assume that the observing cadence and per-epoch duration of pulsar observations remain unchanged across the two 4.5-year time periods, and do not add any new pulsars to the array for the additional 4.5 years of simulated observations.

\subsection{Continuous-Wave Detection}\label{subsec:cw_det}

The \textit{optimal signal-to-noise ratio} is the signal-to-noise ratio that occurs if the signal, $\textbf{\textit{s}}$, is optimally matched to the filter $\textit{s}$, such that $\textbf{\textit{s}} = \textit{s}$ \citep{jaranowski_data_1998}.
The signal, $\textbf{\textit{s}}$, is the contribution from an individual binary, given by Equation \ref{eq:signal_i}.
Following \citet{gardiner_characterizing_2025_fixed}, we assume that our continuous-wave signal follows the same assumptions as our filter model, i.e.,  $\textbf{\textit{s}} = \textit{s}$.

We model the stochastic background as a common-uncorrelated red noise process, rather than as a cross-correlated process in \texttt{ENTERPRISE}, before calculating the optimal statistic for Hellings-Downs correlations. 
This approximation allows us to separate the noise covariance matrix into per-pulsar contributions, without cross-correlated terms, significantly simplifying the matrix inversion process.

The optimal signal-to-noise ratio for a network of pulsars is the square root of the sum of each pulsar's $(\mathrm{S/N})_{\mathrm{CW}, I}$ squared in the array
\begin{equation} \label{eq:opt_SNR}
    (\mathrm{S/N})_{\mathrm{CW}} = \sqrt{\sum_{I}^{N_{\mathrm{pulsars}}} \left((\mathrm{S/N})_{\mathrm{CW},{I}}\right)^2}.
\end{equation}

We use the methods in \texttt{ENTERPRISE} and \texttt{enterprise\_extensions} to perform this calculation for the louder binaries in each population.

We include the noise from the stochastic background.
This has a greater effect at lower frequencies, due to the higher number of binaries there, as in Eq.~\ref{eq:f_prob}.
Although GW signals become less numerous at higher frequencies, they are also easier to detect due to the falling red noise from both the stochastic background and intrinsic pulsar processes.

In \citet{gardiner_characterizing_2025_fixed}, they find that for $(\mathrm{S/N})_{\mathrm{CW}} \geq 5.94$ ($\gtrsim 13$), 50\%  (100\%) of their injected sources are recovered successfully.
At $(\mathrm{S/N})_{\mathrm{CW}} = 5.94$ the false alarm probability is at 5\%, and the false dismissal probability is at $\approx 14\%$.
We thus use $(\mathrm{S/N})_{\mathrm{CW}} = 5.94$ as a conservative threshold for continuous-wave detection in our analysis following this work.

\begin{figure}[htbp]
    \centering
    \includegraphics[width=\linewidth]{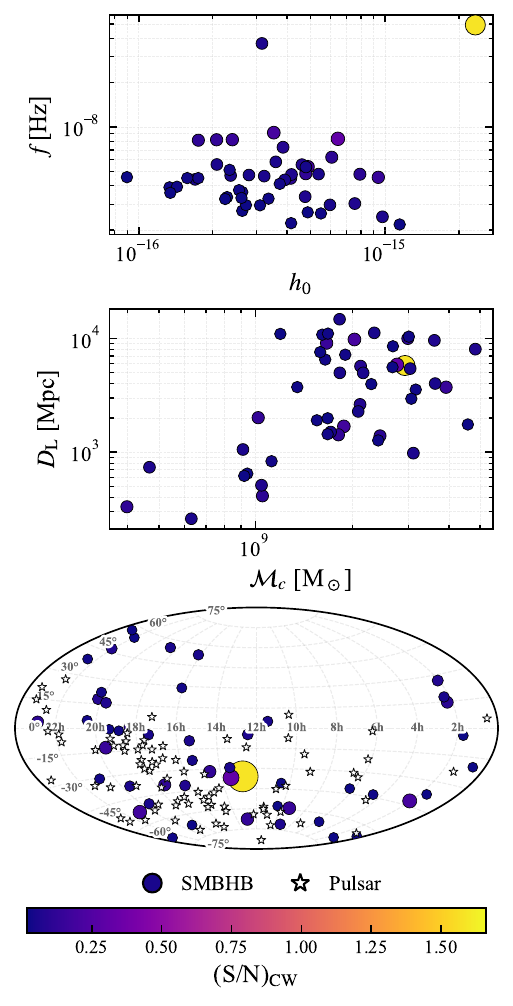}
    \caption{Binary parameters of the SMBHB systems with the highest continuous-wave $(\mathrm{S/N})_{\mathrm{CW}}$, as shown by the color bar, in an individual intermediate scenario simulation of an MPTA 9.0-year dataset. \textit{Top}: dimensionless strain amplitude and gravitational-wave frequency in Hertz, \textit{middle}: chirp mass in solar masses and luminosity distance in megaparsecs, and \textit{bottom}: sky location of the binaries (colored circles, where color and circle size correspond to $(\mathrm{S/N})_{\mathrm{CW}}$), overlaid with the MPTA 4.5-year data release pulsars (white stars).}
    \label{fig:cgw_realistic_params}
\end{figure}

For one simulation of an MPTA 9.0-year data set with the intermediate population scenario, we show the 100 loudest continuous-wave sources that were injected and their properties in Fig.~
\ref{fig:cgw_realistic_params}.
The binaries with the highest amplitude, $h_0$, tend to be the binaries with the highest continuous-wave $(\mathrm{S/N})_{\mathrm{CW}}$.
This amplitude is generally dominated by higher chirp mass, as shown in the second panel.
The gravitational-wave frequency tends to be clustered around $6-25$ nHz, where the MPTA is most sensitive (see Figure \ref{fig:sensitivity_curve} and Section \ref{sec:PTA_opt}).
The distance of the binary does not dramatically influence the loudest binaries in this simulation.
In this case, we find that most resolvable binaries tend to lie in the southern hemisphere, where most MPTA pulsars are clustered.
This is due to the antenna pattern function shown in Equation \ref{eq:ant_pat}.
The greater the separation in the sky between a pulsar and an SMBHB, the lower the antenna pattern function for that source, translating to a lower $(\mathrm{S/N})_{\mathrm{CW}}$.

\section{Results}\label{sec:resolved}

\subsection{Nearest Binary}
\label{subsec:nearest}
\begin{figure}[htbp]
    \centering
    \includegraphics[width=\linewidth]{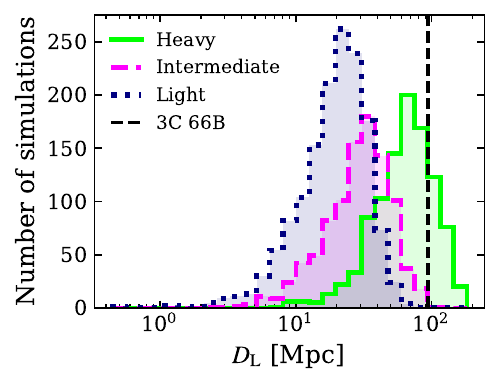}
    \caption{Distribution of the nearest binary from each of the light (dotted navy), intermediate (dashed magenta), and heavy (solid lime) scenarios for forming nearby SMBHB systems through the population synthesis approach developed. Each y-axis value represents the nearest SMBHB within one simulation run. The dashed vertical line at 94 Mpc shows the luminosity distance of the nearest electromagnetic SMBHB candidate, 3C~66B \citep{van_den_bosch_hunting_2015}.}
    \label{fig:nearest_binary}
\end{figure}

In each synthesized population, we determine the nearest simulated SMBHB.
This allows us to compare with nearby SMBHB candidates from electromagnetic observations.
Figure \ref{fig:nearest_binary} shows the distributions of the nearest SMBHB within each simulation for the light (dotted navy), intermediate (dashed magenta), and heavy (solid lime) scenarios.

The distribution peaks at luminosity distances of approximately $20$, $30$, and $80\,\mathrm{Mpc}$ for the light, intermediate, and heavy scenarios, respectively.
The light model favors less massive binaries compared to the intermediate or heavy models; it therefore requires a larger number of systems to reproduce the stochastic background as measured by pulsar timing arrays. 
As a consequence, it is more likely that the light model contains a nearby binary than the intermediate or heavy models. 
In comparison to our populations, the nearest electromagnetic SMBHB candidate is 3C~66B (94 Mpc away from Earth; \citealt{van_den_bosch_hunting_2015}).
All scenarios have SMBHB systems closer than 3C~66B, indicating that our synthesized populations produce sufficiently close binaries to be consistent with current electromagnetic constraints.

\subsection{Loudest Binary}
\label{subsec:loudest}
Early studies predicted that the first continuous-wave source detectable by PTAs would have a gravitational-wave frequency $\sim 3-10$ nHz (\citealt{kelley_single_2018}), a chirp mass of $\gtrsim 10^{8} \; \rm{M}_{\odot}$, and be relatively close-by ($z \lesssim 1.5$; \citealt{sesana_measuring_2010, rosado_expected_2015, mingarelli_local_2017}).
\citet{sesana_gravitational_2009} suggested that detecting such sources would most likely require facilities such as the future Square Kilometre Array that are more sensitive than most current radio telescopes used in PTAs.
Consistent with this expectation, \citet{becsy_exploring_2022_fixed} found that when forecasting for a NANOGrav 15-year data set, only 6\% of their realizations using semi-analytical SMBHB models coupled with the Illustris simulation results had an SMBHB with S/N $> 5.0$.

Evidence for a stochastic gravitational-wave background with an amplitude that is larger than expected from theoretical expectations (e.g., \citealt{rajagopal_ultra--low-frequency_1995, dvorkin_nightmare_2017}) has increased the prospect of detecting individual continuous-wave sources with current PTA data sets, as it may imply a more numerous and/or more massive SMBHB population \citep[e.g.,][]{agazie_nanograv_2023}.
\citet{gardiner_characterizing_2025_fixed} used results from \citet{agazie_nanograv_2023} to include constraints from the stochastic background to update previous forecasts from \citet{becsy_exploring_2022_fixed, becsy_how_2023_fixed}.
They found that detectable continuous-wave sources are expected to have frequencies of $4-12$ nHz, chirp masses between $0.7-20\times 10^{9} \;\rm{M}_{\odot}$, and luminosity distances of $60 \; \rm{Mpc} - 8 \; \rm{Gpc}$, broadly consistent with early predictions.
In contrast to the 6\% estimate of \citet{becsy_exploring_2022_fixed}, 53\% of their SMBHB populations yielded continuous-wave detections in the current data.
More recently, \citet{agarwal_nanograv_2026} reported two potential (Bayes factor greater than one) continuous-wave candidates.

\begin{figure}[htbp]
    \centering
    \includegraphics[width=\linewidth]{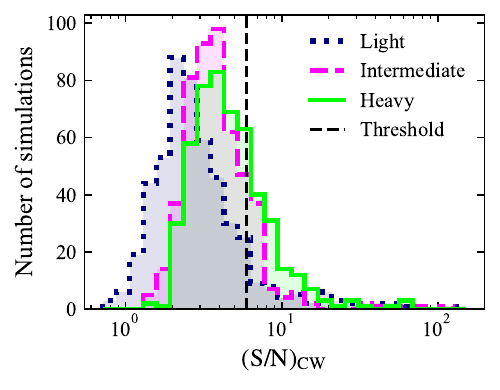}
    \caption{Distributions for the binary with the highest continuous-wave signal-to-noise ratio, $(\rm S/N)_{\rm CW}$, for a lighter, more numerous SMBHB population (dotted navy), an intermediate population (dashed magenta), and a heavier, less numerous SMBHB population (solid lime). A detection threshold of $(\mathrm{S/N})_{\mathrm{CW}}$ $= 5.94$ is shown in dashed black. }
    \label{fig:loudest_binary_SNR}
\end{figure}

Motivated by previous studies of the NANOGrav 15-year data set, we perform a similar analysis to \citet{gardiner_characterizing_2025_fixed}, incorporating information about the stochastic background from the MeerKAT 4.5-yr data set into cosmological populations of SMBHB systems. 
For each population, we calculate the loudest binary using Eq.~\ref{eq:opt_SNR}.
Fig.~\ref{fig:loudest_binary_SNR} shows the distributions of the continuous-wave signal-to-noise ratio ($(\mathrm{S/N})_{\mathrm{CW}}$) for the loudest binary in each simulation, following the color scheme from Fig.~\ref{fig:nearest_binary}.
We find that for heavier SMBHB distributions for our populations, $(\mathrm{S/N})_{\mathrm{CW}}$ increases.
As the signal is proportional to $\mathcal{M}_c^{5/3}$, increasing the mass of the binaries through a higher $m_{\rm{char}}$ increases $(\mathrm{S/N})_{\mathrm{CW}}$.

The dashed black line at $(\mathrm{S/N})_{\mathrm{CW}}=5.94$ in Fig.~\ref{fig:loudest_binary_SNR} is our threshold for continuous-wave detection in a PTA (see discussion in Section \ref{subsec:cw_det}).
We find that the majority of the highest $(\mathrm{S/N})_{\mathrm{CW}}$ SMBHB systems in all scenarios fall below this detection threshold, with detection fractions of $10\%$, $16\%$, and $27\%$ for the light, intermediate, and heavy scenarios, respectively.
Given the absence of a confirmed continuous-wave detection to date, these fractions are not unexpected.
By comparison, \citet{gardiner_characterizing_2025_fixed}, found that 53\% of their populations conditioned on the the stochastic background have a detectable source.
This difference may partly reflect the data sets used: \citet{gardiner_characterizing_2025_fixed} uses the NANOGrav 15-yr data set, for which the stochastic-background significance was $(\mathrm{S/N})_{\mathrm{SGWB}}=3.5-4.0\sigma$, compared to $(\mathrm{S/N})_{\mathrm{SGWB}}=3.0-3.4\sigma$ for the MPTA 4.5-year data set and its 9.0-year forecast used in this analysis.

\begin{figure*} [htbp]
 \begin{subfigure}{0.49\textwidth}
     \includegraphics[width=\textwidth]{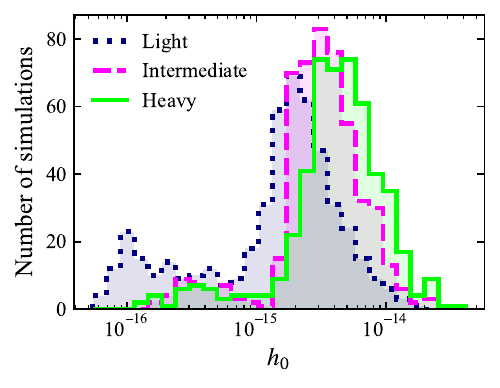}
     \label{fig:h0}
 \end{subfigure}
 \hfill
 \begin{subfigure}{0.49\textwidth}
     \includegraphics[width=\textwidth]{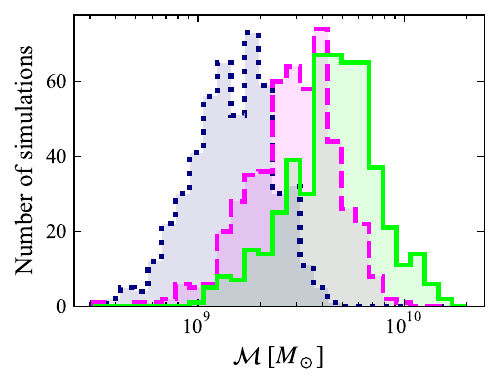}
     \label{fig:Mc}
 \end{subfigure}
 
 \vspace{-6pt}
 \begin{subfigure}{0.49\textwidth}
     \includegraphics[width=\textwidth]{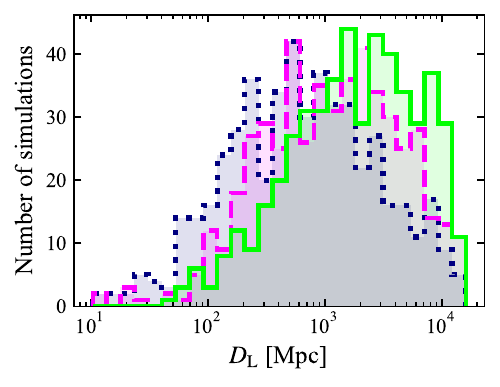}
     \label{fig:D_comov}
 \end{subfigure}
 \hfill
 \begin{subfigure}{0.49\textwidth}
     \includegraphics[width=\textwidth]{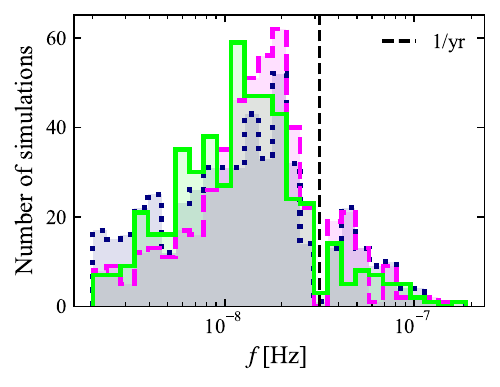}
     \label{fig:f}
 \end{subfigure}
  \caption{Distributions for the parameters of the binary with the highest continuous gravitational-wave signal-to-noise ratio for a lighter, more numerous SMBHB population (dotted navy), an intermediate SMBHB population (dashed magenta), and a heavier, less numerous SMBHB population (solid lime). \textit{Top left}: dimensionless strain amplitude, \textit{top right}: chirp mass in solar masses, \textit{bottom left}: luminosity distance in megaparsecs, and \textit{bottom right}: gravitational-wave frequency in Hertz. The dashed black line in the bottom right panel corresponds to a frequency of $yr^{-1}$, where a dearth of SMBHB systems at this frequency is due to the timing model in \texttt{ENTERPRISE}.}
 \label{fig:loudest_binary}
\end{figure*}

In Fig.~\ref{fig:loudest_binary}, we show the amplitude ($h_0$), the chirp mass ($\mathcal{M}_c$), the luminosity distance ($D_{\rm{L}}$), and the gravitational-wave frequency ($f$), for the loudest binary in each simulation. 
We find that the amplitude of the highest $(\mathrm{S/N})_{\mathrm{CW}}$ binary increases as we move to mass distributions that allow higher-mass binaries, consistent with the mass dependence of the continuous-wave signal.
We find that for all scenarios, there is an apparent bimodality in the distribution for $h_0$, for one subset with $10^{-15}\lesssim h_0 \lesssim 3\times 10^{-14}$, and another subset with $5\times 10^{-17}\lesssim h_0 \lesssim \times 10^{-15}$.
We discuss this bimodality in more detail in Section \ref{subsubsec:bimodality}.
The chirp mass distribution of the loudest binary follows from the mass distributions for each scenario: less massive black holes are likely to be drawn for the light case, and more massive for the heavy case.
The luminosity-distance distribution of the loudest binary peaks at $\sim 500$ Mpc for the light population, $\sim 1500$ Mpc for intermediate, and $\sim 3000$ Mpc for heavy.
Since the simulations are truncated at $z=2$, these results suggest that the first resolvable CW source could occur at a significant redshift; however, it is likely to occur closer as the distributions all peak before the maximum allowed luminosity distance.
In our population synthesis, we assume that the sources are uniformly distributed in comoving volume. 
However, this assumption is unlikely to remain valid at high redshift since at $z=2$, the Universe was approximately $3$ Gyr old, and the SMBHB population is expected to evolve significantly with redshift (e.g., \citealt{chen_constraining_2019, kozhikkal_mass-redshift_2024}).

The $95\%$ credible intervals for the gravitational-wave frequency of the loudest binary are $2.3$--$77\,\mathrm{nHz}$, $2.6$--$74\,\mathrm{nHz}$, and $2.8$--$73\,\mathrm{nHz}$ for the light, intermediate, and heavy scenarios, respectively. 
In all cases, the distribution peaks at approximately $15$--$20\,\mathrm{nHz}$. 
The dashed black line corresponds to a frequency of $\text{yr}^{-1}$, where the sensitivity is reduced due to Earth's motion around the Sun.
The frequency peaks lie above the $\sim6\,\mathrm{nHz}$ region of greatest sensitivity for the 9.0-year MPTA data set (Figure~\ref{fig:sensitivity_curve}; Section~\ref{sec:PTA_opt}), where the pulsar red noise and the stochastic background are weaker. 
At higher frequencies, the data becomes dominated by white noise. 
Additionally, because the binary frequency distribution follows $f^{-11/3}$, high-frequency SMBHB systems are intrinsically less common, reducing confusion between SMBHB sources.

\begin{figure}[htbp]
    \centering
    \includegraphics[width=\linewidth]{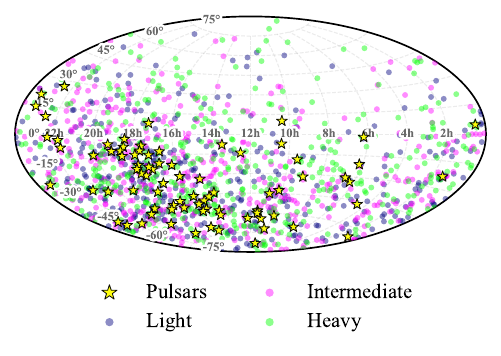}
    \caption{Sky map for the location of the binary in each simulation with the highest continuous gravitational-wave signal-to-noise ratio for a lighter, more numerous SMBHB population (navy), an intermediate SMBHB population (magenta), and a heavier, less numerous SMBHB population (lime). The yellow stars are the locations of MPTA's pulsars used in the 4.5-year analysis.}
    \label{fig:loudest_binary_skymap}
\end{figure}

We show the sky location of the loudest SMBHB in each simulation in Fig.~\ref{fig:loudest_binary_skymap}.
The binaries with the highest continuous-wave signal-to-noise ratio are preferentially located in the southern hemisphere, where the MeerKAT pulsars are most densely distributed, an observational bias due to the MPTA being located in the southern hemisphere.
As discussed in Section~\ref{subsec:cw_det}, this directional dependence arises from the pulsar antenna-pattern function in Equation~\ref{eq:ant_pat}.

\begin{figure}[htbp]
    \centering
    \includegraphics[width=\linewidth]{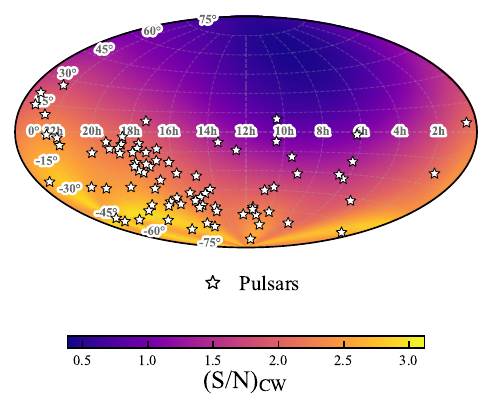}
    \caption{Sky map of the continuous-wave optimal signal-to-noise ratio, $(\mathrm{S/N})_{\mathrm{CW}}$, for a binary with $f = 10$ nHz, $\mathcal{M} = 10^9\,\mathrm{M}_{\odot}$, and $q = 0.5$ at different sky locations in a simulated MPTA 9.0-year data set. The color of the tiling corresponds to the continuous-wave $(\mathrm{S/N})_{\mathrm{CW}}$ of the source located at that sky position, and the white stars show the location of the MPTA's pulsars used in the 4.5-year analysis.}
    \label{fig:skymap_comp}
\end{figure}

Figure~\ref{fig:skymap_comp} shows $(\mathrm{S/N})_\mathrm{CW}$ as a function of sky location for a binary with $f = 10$ nHz, $\mathcal{M} = 10^9\,\mathrm{M}_{\odot}$, and $q = 0.5$ in the simulated MPTA 9.0-year data set. 
Tests with different binary parameters yielded similar sky-location trends. 
The continuous-wave signal-to-noise ratio at the most sensitive sky location is approximately six times larger than that at the least sensitive location. 
The most sensitive region lies near the Galactic plane, reflecting the pulsar distribution.
However, Galactic extinction and source confusion may hinder electromagnetic identification of binaries in this region, potentially limiting opportunities for multi-messenger observations.

\begin{figure*}[htbp] \label{fig:binary_high_CGW_thresh}
 \begin{subfigure}{0.49\textwidth}
     \includegraphics[width=\textwidth]{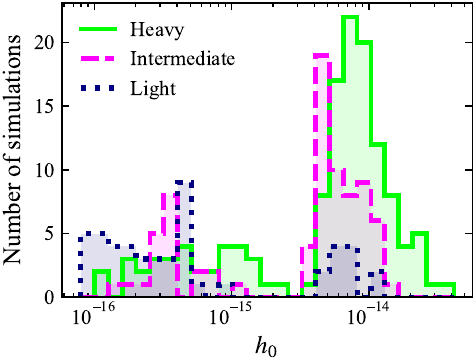}
     \label{fig:ah0_thresh}
 \end{subfigure}
 \hfill
 \begin{subfigure}{0.49\textwidth}
     \includegraphics[width=\textwidth]{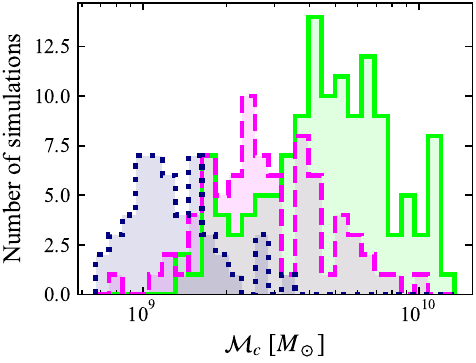}
     \label{fig:Mc_thresh}
 \end{subfigure}
 
 \vspace{-6pt}
 \begin{subfigure}{0.49\textwidth}
     \includegraphics[width=\textwidth]{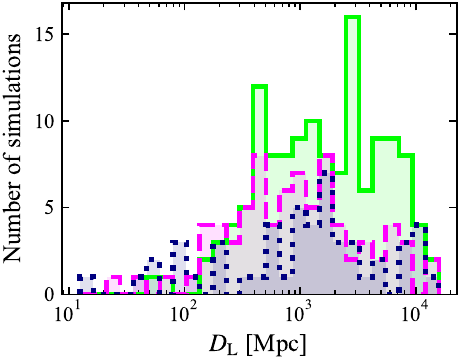}
     \label{fig:D_comov_thresh}
 \end{subfigure}
 \hfill
 \begin{subfigure}{0.49\textwidth}
     \includegraphics[width=\textwidth]{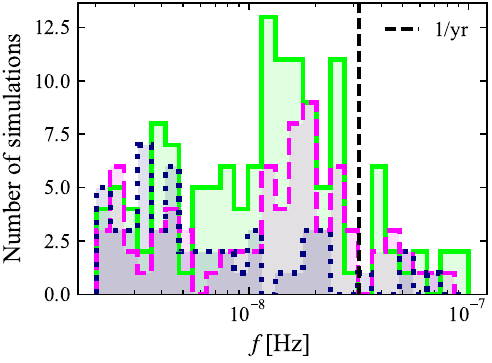}
     \label{fig:f_thresh}
 \end{subfigure}
  \caption{As Figure \ref{fig:loudest_binary}, except with restricted distributions to only include binaries that exceed the detection threshold of $(\mathrm{S/N})_{\mathrm{CW}}$ $= 5.94$.}
 \label{fig:loudest_binary_threshold}
\end{figure*}

Figure \ref{fig:loudest_binary_threshold} shows the same distributions as Figure \ref{fig:loudest_binary}, but here we have restricted the distributions to include only binaries that exceed the detection threshold, $(\mathrm{S/N})_{\mathrm{CW}}$ $= 5.94$.
This selection significantly reduces the number of sources.
The amplitude and chirp-mass distributions are broadly consistent with those of the full population (Fig.~\ref{fig:loudest_binary}). 
The bimodality in the amplitude is still present in the detectable binary population, which we discuss further on.
The luminosity distance distributions show more scatter than the full population.
There is also a shift towards lower distances compared to Fig. \ref{fig:loudest_binary}, as expected for detectable sources.

The 95\% credible interval for the GW frequency of detected binaries that exceed the detection threshold is $2.1$--$51\,\mathrm{nHz}$, $2.2$--$58\,\mathrm{nHz}$, and $2.3$--$74\,\mathrm{nHz}$ for the light, intermediate, and heavy scenarios, respectively.
These intervals are generally similar to the frequencies of the all-binary populations shown in Figure \ref{fig:loudest_binary}, albeit there are few sources with $f \gtrsim 80$ nHz in the detected-binary populations for all scenarios.
This arises as the forecasted MPTA 9.0-year data set is most sensitive at $6$ nHz, with decreasing sensitivity at higher frequencies.

\subsubsection{Strain amplitude bimodality} \label{subsubsec:bimodality}
\begin{figure*}[htbp]
    \centering
    \includegraphics[width=\textwidth]{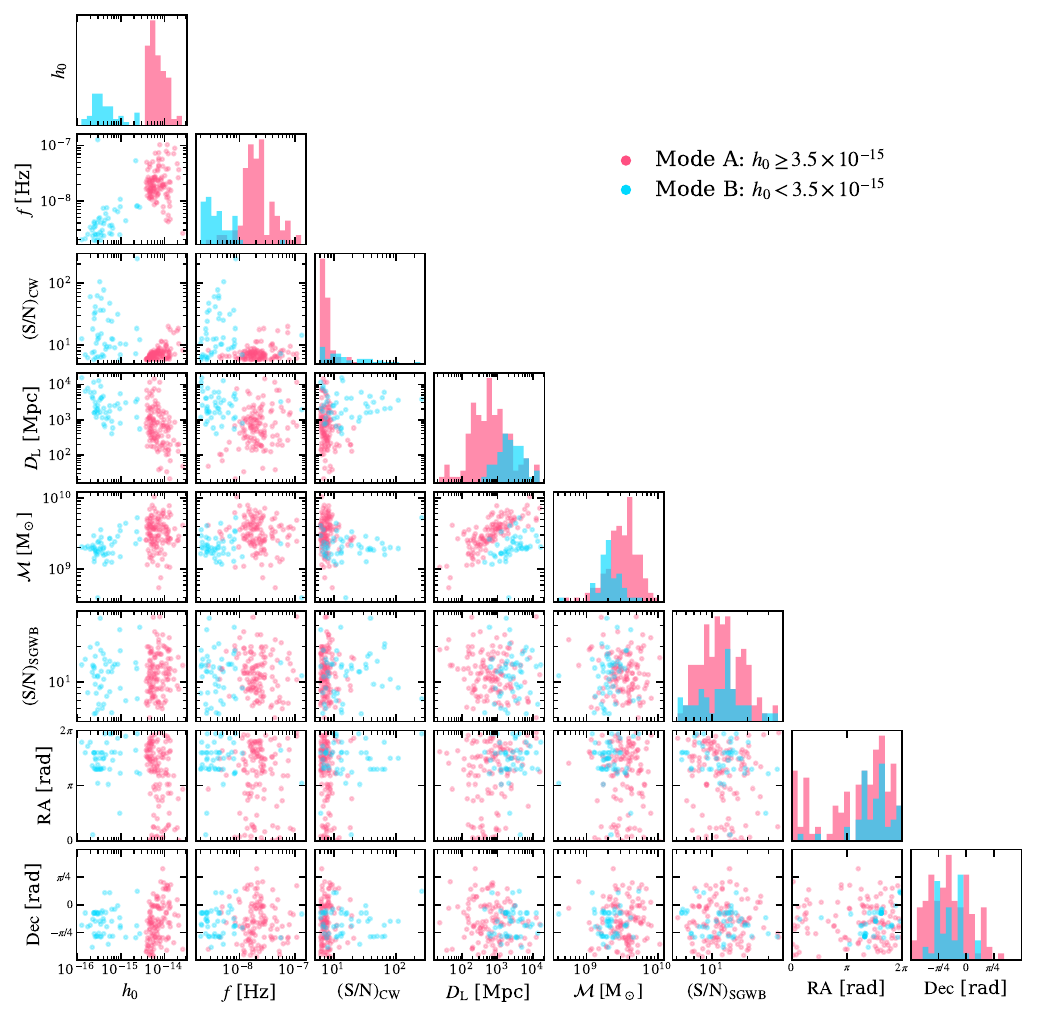}
    \caption{Parameter distributions for the detectable binaries in the intermediate scenario, divided into two sub-populations: mode A) $h_0 \geq 3.5\times 10^{-15}$ (pink), and mode B) $h_0 < 3.5\times 10^{-15}$ (cyan). This threshold was chosen from visual inspection of the bimodality in the $h_0$ distribution, in order to analyze the correlations in these two sub-populations.}
    \label{fig:corner_bimodality}
\end{figure*}

In both the detectable- and all-binary populations for all scenarios, we find that $h_0$ is bimodal, with a subset of the populations with $10^{-15}\lesssim h_0 \lesssim 3\times 10^{-14}$, and a subset with $5\times 10^{-17}\lesssim h_0 \lesssim \times 10^{-15}$.
To examine this further, Fig~\ref{fig:corner_bimodality} shows the joint parameter distributions for the detectable binaries in the intermediate scenario, categorized into two sub-populations: i) $h_0 < 3.5\times 10^{-15}$ (cyan), and ii) $h_0 \geq 3.5\times 10^{-15}$ (pink).
We choose this $h_0$ threshold from visual inspection of the bimodality in the data.
We categorize these as two modes:
\begin{itemize}
    \item \textbf{Mode A}: Higher $h_0$, higher chirp mass, higher frequency, broad distribution in luminosity distance, weaker $(\mathrm{S/N})_{\mathrm{CW}}$, broad distribution of sky location (right ascension [RA] and declination [Dec]).
    \item \textbf{Mode B}: Lower $h_0$, lower chirp mass, lower frequency, farther luminosity distance, higher $(\mathrm{S/N})_{\mathrm{CW}}$, tightly clustered right ascension and declination near the most sensitive sky location for the MPTA.
\end{itemize}
The distributions for the chirp mass and distance are correlated with the amplitude, as expected, given that $h_0\propto \mathcal{M}^{5/3}/D_{L}$.
For the B mode, this explains the lower chirp mass and farther distances than the A mode for the difference in amplitude.
The clustering of the sky location (RA and Dec) of the B mode as compared to the broad distribution for the A mode indicates that the sky location dependence is likely one of the causes for these two modes.
Following from Fig.~\ref{fig:skymap_comp}, tight clustering around the most sensitive sky location can result in a factor of six greater $(\mathrm{S/N})_{\mathrm{CW}}$ than in the least sensitive sky location.
The lower frequency of the mode B population may be because the simulated MPTA 9.0-year data set is most sensitive at $\sim 6$ nHz (see Fig~\ref{fig:sensitivity_curve}).
Although the stochastic background noise may be higher at low frequencies due to the higher number of binaries, the increased sensitivity at these frequencies may compensate such that the comparatively lower $h_0$ can lead to a high $(\mathrm{S/N})_{\mathrm{CW}}$.

We find that all of the highest $(\mathrm{S/N})_{\mathrm{CW}}$ sources are from the B mode, despite their lower $h_0$.
This is an unexpected result, as the continuous-wave signal is proportional to $h_0/(2\pi f)$, so higher $h_0$ sources would be expected to have the highest $(\mathrm{S/N})_{\mathrm{CW}}$.
However, the sky location and frequency sensitivity of the PTA affect the $(\mathrm{S/N})_{\mathrm{CW}}$, and may lead to these results.
A key effect we identify from this is that intrinsically loud sources, from high mass and proximity, are not the only potentially resolvable sources; sources that can influence the detector response substantially from frequency and sky location, for example, can also emerge as loud, detectable sources.
Understanding the origin of these two populations is saved for future study.

\subsubsection{Per-Pulsar Contributions} \label{sec:pulsar_cont}
The optimal continuous-wave signal-to-noise ratio,  $(\mathrm{S/N})_{\mathrm{CW}}$, increases with the duration of observations and with reduced timing residual noise (\citealt{burt_optimizing_2011, hazboun_realistic_2019}).

\begin{figure}[htbp]
    \includegraphics{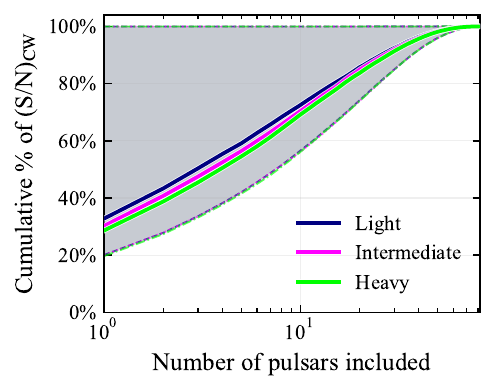}
    \centering
    \caption{Cumulative fraction of $(\mathrm{S/N})_{\mathrm{CW}}$ as a function of decreasing pulsar importance for the light (navy), intermediate (magenta), and heavy (lime) scenarios. The shaded regions enclosed by the corresponding colored dot-dashed lines denote the 95\% credible interval for the relevant scenario.}
    \label{fig:psr_cont}
\end{figure}

In Fig.~\ref{fig:psr_cont}, we show how much each pulsar contributes to $(\mathrm{S/N})_{\mathrm{CW}}$ using Eq. \ref{eq:opt_SNR} for the light (navy), intermediate (magenta), and heavy (lime) scenarios, where the colored shaded regions correspond to the respective 95\% credible intervals.
Typically $\sim$ 70\% of the signal comes from the ten most sensitive pulsars.
These important pulsars generally have the lowest levels of noise, leading them to contribute the most to the $(\mathrm{S/N})_{\mathrm{CW}}$.
For example, across all scenarios, the pulsars J2241--5236, J1909--3744, J1946--5403 and J2129--5721 contribute to the majority of the total continuous-wave signal.
These are among the most precisely timed pulsars in the array and have comparatively low noise levels, particularly J2241--5236 and J1909--3744.
They also have similar observing baselines and cadences to all other pulsars in the array, indicating that the dominant contributions reflect the timing precision.

To further increase $(\mathrm{S/N})_{\mathrm{CW}}$ and maximize the detection probability, we focus on the ten most sensitive pulsars when changing observation strategies to maximize the returns from increased telescope time.

\section{Pulsar Timing Optimization} \label{sec:PTA_opt}
Detection of a stochastic GW background requires at least 40--50 pulsars with broad sky coverage to enable the determination of the Hellings-Downs correlation (\citealt{jenet_detecting_2005, siemens_stochastic_2013}).
In contrast, detection of individual continuous-wave sources is generally improved by increasing the observing cadence of the most precisely timed pulsars in the array, as these pulsars make the largest contribution to the detection signal-to-noise ratio.
Adding pulsars to a timing array can improve sensitivity for continuous-wave source detection.
However, this improvement depends on their sky distribution. 
In particular, sensitivity to individual continuous-wave sources improves when the additional pulsars are clustered closer to the primary pulsars in the array, contrary to stochastic background searches, for which broad sky coverage is essential (\citealt{burt_optimizing_2011}).
For a fixed observing time, introducing pulsars with higher noise can decrease the likelihood of continuous-wave detection by decreasing the observing time on the best-timed pulsars  (\citealt{lee_optimal_2012}).
Here we focus on modifications to the observing strategy that can enhance the detectability of individual continuous-wave sources.

\citet{christy_optimization_2014} found that maximizing observation time on the highest-precision pulsars can improve sensitivity to individual SMBHB systems by factors of $1.5-4.0$, depending on pulsar noise.
They further noted that adding pulsars in directions with an enhanced prior probability of hosting an SMBHB system, e.g., towards the Virgo Cluster, may improve the sensitivity of the array to these sources.
\citet{baier_sensitivity_2025} simulated populations of SMBHB systems and found that re-allocating observation time from poorly timed pulsars to high-precision pulsars increases the detection probability of continuous-wave sources. 
For example, they found that stopping observations of the ten least sensitive pulsars and increasing the cadence for the most sensitive pulsars resulted in a $\sim20\%$ ($\sim50\%$) increase in the detectable volume at 20 nHz (50 nHz) for an IPTA-like detector, with minimal effects at lower frequencies.
We aim to complement this study by considering SMBHB populations that are consistent with current PTA observation constraints on the stochastic background. 
More recently, \citet{gitika_optimising_2025} found that increasing the observing cadence of the highest-precision pulsars optimizes continuous-wave sensitivity in the MeerKAT PTA, and only marginally decreases sensitivity to the stochastic background.
Specifically, a fourfold increase in the cadence of observations for the best 10 pulsars over three years, following five years of observations, resulted in a 14\% increase in sensitivity to continuous-wave sources relative to the fiducial eight-year observing plan.

\subsection{Implementing Changes to the Pulsar Timing Strategy} \label{subsec:changes_PTA}
Most of the loudest and most detectable sources in our simulations predominantly lie in the $\sim 2-70$ nHz frequency range.
Improving the PTA sensitivity in this band will increase the $(\mathrm{S/N})_{\mathrm{CW}}$ of detectable sources and hence the detectable fraction. 
At higher frequencies, the lower number of binaries found at these frequencies also reduces source-confusion effects between binaries.

\begin{figure}[htbp]
    \centering
    \includegraphics[width=\linewidth]{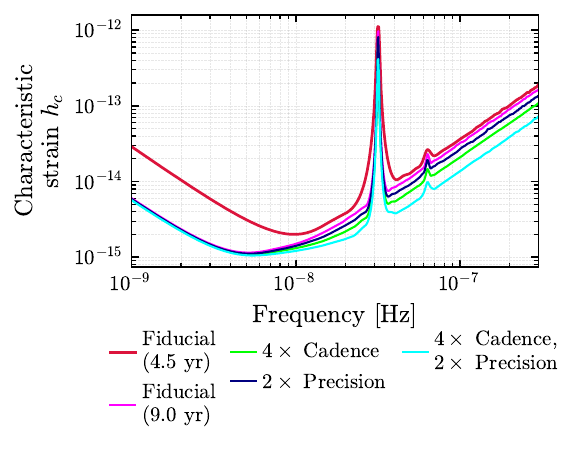}
    \vspace{-20pt}
    \caption{Sensitivity curves using \texttt{hasasia} for the MPTA 4.5-year data release pulsars (crimson), as well as this data combined with another 4.5 years of data with: the current observing plan (magenta), the most sensitive ten pulsars with an increase in cadence by a factor of four and otherwise unchanged (lime), the most sensitive ten pulsars with an increase in precision by a factor of two and otherwise unchanged  (navy), and a combination of these latter two (cyan). These curves are made without considering dispersion measure and chromatic noise for simplicity. The characteristic strain is defined as $h_c = A \left(f/\rm{yr}^{-1} \right)^{-2/3}$, where $A$ is from Equation \ref{eq:PSD}.}
    \label{fig:sensitivity_curve}
\end{figure}

Using \texttt{hasasia}, Fig.~\ref{fig:sensitivity_curve} shows the sensitivity of the MPTA to a continuous-wave source for the MPTA 4.5-year data release, and then forecasted with various observation strategies to a 9.0-year dataset, ignoring pulsar chromatic noise processes for simplicity.
We find that improving the timing precision of the ten most sensitive pulsars provides a factor of $\sim 1.2$ sensitivity increase at the target frequencies.
Comparatively, increasing the cadence of these pulsars by a factor of four results in a sensitivity increase by a factor of $\sim1.5$.
Combining both approaches results in an overall sensitivity increase by a factor of $\sim 2.1$ relative to the current MPTA observing strategy.
The MPTA 4.5-year sensitivity curve is most sensitive near $\sim 10$ nHz. 
Extending the observing baseline to 9 years shifts the peak sensitivity to a lower frequency, $\sim 6$ nHz.
At frequencies $\lesssim 6$ nHz, all of the forecasted pulsar timing array sensitivity curves are similar, while at higher frequencies, all of the altered observing strategies improve compared to the fiducial observing plan.
These improvements may increase the $(\mathrm{S/N})_{\mathrm{CW}}$ of many sources that were below the threshold in the fiducial MPTA observing plan.

Given that the pulsars with the highest timing precision dominate the continuous-wave $(\mathrm{S/N})_{\mathrm{CW}}$ (Section \ref{sec:pulsar_cont}), we apply the precision and cadence improvements to the ten most sensitive pulsars.
We have not conserved total telescope observation time in this approach; however, we implemented this in some observing strategies by decreasing the cadence of the poorly timed pulsars.
The MPTA 4.5-year pulsars are currently observed once every $\sim 14$ days, such that an increase by a factor of four requires observations every $3-4$ days.
Such high-cadence observations for a subset of pulsars are feasible. 
For example, the Canadian Hydrogen Intensity Mapping Experiment telescope (CHIME) has monitored several NANOGrav pulsars at near-daily cadence \citep{agazie_chime-o-grav_2026}.
Improving timing precision will generally increase the observation time per pulsar visit.
Although wideband receivers can significantly improve timing precision for pulsars not dominated by jitter noise \citep{lam_optimizing_2018}, several of most sensitive pulsars are already limited by jitter. 
Since the arrival-time uncertainty is inversely proportional to the number of integrated pulses \citep{shannon_pulse_2012}, improving the timing precision by a factor of two requires approximately four times the integration time.
We use this solely as a case study, but in the future, with instruments such as the Square Kilometre Array, further sensitivity increases may be possible compared with current data sets.

\subsection{Injecting SMBHB signals into optimized PTAs}
We examine how these changes affect the $(\mathrm{S/N})_{\mathrm{CW}}$ of our SMBHB sources by augmenting the additional 4.5-year simulated datasets with more observations for higher cadence and an increased duration for higher precision. 
We then follow the same continuous-wave analysis pipeline as for the fiducial MPTA 9.0-year PTA, except with the new observation methods. 
We consider the following observing strategies:
\begin{itemize}
    \item \textbf{Fiducial:} Following the MPTA 4.5-year observing strategy, we adopt a cadence of approximately one observation every 1/(2 weeks) and integration times of 256--2048\,s depending on the pulsar \citep{miles_meerkat_2024}. Here the minimum duration is used when the required precision is achieved.
    \item \textbf{4 $\times$ Cadence:} The same as fiducial, except the cadence of the ten most sensitive pulsars (as discussed in Section \ref{sec:pulsar_cont}) is increased to $\approx 2/(\mathrm{week)}$.
    \item \textbf{2 $\times$ Precision:} The same as fiducial, except the observation duration of the ten most sensitive pulsars is increased by a factor of four, improving the timing precision by a factor of two as discussed in Section \ref{subsec:changes_PTA}.
    \item \textbf{Combined:} Combining the increased cadence and precision cases above, such that the ten most sensitive pulsars are observed twice a week and for four times as long compared to the fiducial case, while the remaining pulsars follow the fiducial strategy.
    \item \textbf{(TT Cons.):} Denotes that the telescope time across all pulsars for an observing strategy is the same as in the fiducial case. When increasing the number/duration of observations of the most sensitive pulsars, fewer observations are conducted on the remaining pulsars.
    \item \textbf{Max. Cadence:} The cadence of a selected subset of the most sensitive pulsars is maximized by allocating the maximum number of observations to the most sensitive pulsars while reducing observations of the remaining pulsars to a cadence of $1/(3\;\mathrm{months})$, conserving the total telescope time.
\end{itemize}

\begin{figure}[htbp]
 \begin{subfigure}{0.92\columnwidth}
    \vspace{-10pt}
     \includegraphics[width=\columnwidth]{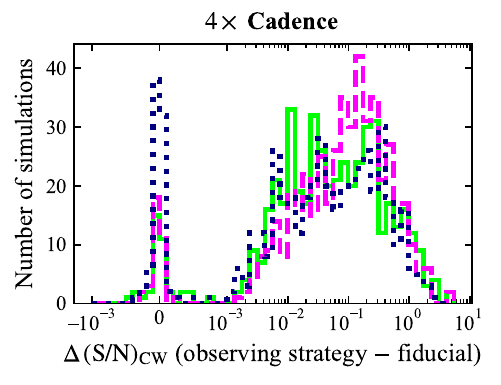}
     \label{fig:5cad_snr}
 \end{subfigure}
 
 \vspace{-30pt}
 \begin{subfigure}{0.92\columnwidth}
     \includegraphics[width=\columnwidth]{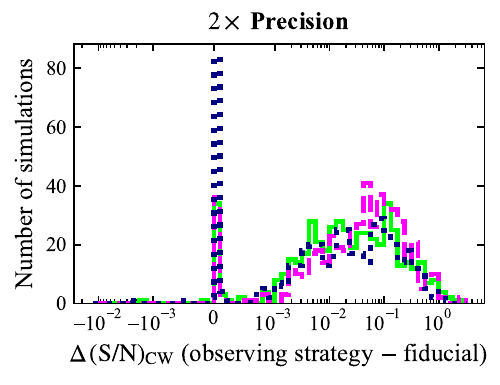}
     \label{fig:4prec_snr}
 \end{subfigure}
 \vspace{-10pt}
 \begin{subfigure}{0.92\columnwidth}
    \vspace{-30pt}
     \includegraphics[width=\columnwidth]{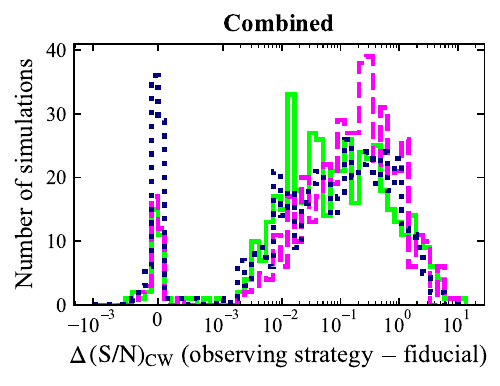}
     \label{fig:5cad_4prec_snr}
 \end{subfigure}
  \caption{Distributions for the change in continuous-wave signal-to-noise $(\rm S/N)_{\rm CW}$ relative to the fiducial MPTA 9.0-year observing strategy of the binary with the highest continuous gravitational-wave optimal signal-to-noise ratio for a lighter, more numerous SMBHB population (dotted navy), an intermediate SMBHB population (dashed magenta) and a heavier, less numerous SMBHB population (solid lime). \textit{Top}: PTA with cadence increased by a factor of four for the ten pulsars contributing most to $(\mathrm{S/N})_{\mathrm{CW}}$, \textit{middle}: PTA with precision increased by a factor of two for the ten pulsars contributing most to the $(\mathrm{S/N})_{\mathrm{CW}}$, \textit{bottom}: PTA with cadence increased by a factor of four and precision increased by a factor of two for the ten pulsars contributing most to $(\mathrm{S/N})_{\mathrm{CW}}$.}
 \label{fig:synth_pta_snr}
\end{figure}

We show the change in continuous-wave $(\mathrm{S/N})_{\mathrm{CW}}$ relative to the fiducial strategy of the loudest binary in each simulation for a factor of four increase in cadence for the ten most sensitive pulsars (middle), a factor of two increase in precision (top), and both of these changes combined (bottom) in Fig.~\ref{fig:synth_pta_snr}.
The peaks at zero for all scenarios and observing strategies show that for some binaries, the change in observing strategy doesn't affect the loudness of some sources --- this likely occurs when the loudest source is low-frequency, such that the sensitivity change is negligible compared to the fiducial case.
We find a general shift to higher $(\mathrm{S/N})_{\mathrm{CW}}$ for all sources in the changed observing strategy datasets, as expected for more sensitive PTAs.
This is most evident in the combined high-precision and high-cadence case, where the change in $(\mathrm{S/N})_{\mathrm{CW}}$ peaks at $\sim 0.1-1.0$ for the three population scenarios.

As PTAs increase sensitivity to gravitational waves at higher frequencies, there are two effects: (1) previously sub-threshold signals may become resolvable, and (2) the previous loudest binary may be superseded by a binary that has become louder in a higher frequency band.
In Table \ref{tab:detection_fraction}, we show the detection fraction for all scenarios and pulsar timing array observing strategies.
We find that each of the changes to the standard MPTA observing plan increases the detection fraction, except when conserving telescope time.

By considering the fraction of simulations with SMBHB systems above the $(\mathrm{S/N})_{\mathrm{CW}} = 5.94$ detection threshold, we find that increasing the precision of the ten most sensitive pulsars by a factor of two increased the detection fraction to 11\% for the light scenario, 17\% for the intermediate, and 28\% for the heavy case.
Increasing the cadence of the ten pulsars chosen increased the detection fraction to 12\%, 18\%, and 30\%, respectively.
The combination of these two increased the detection fraction to 12\%, 21\%, and 33\%, respectively.
When conserving telescope time, for each of higher cadence and higher precision, the detection fraction decreased by $\sim$ a per cent.
For the combined case with telescope time conserved, we find that the detection fraction decreased by up to ten per cent compared with the fiducial observing strategy, indicating the need for observations with many pulsars when trying to detect these binaries.

\begin{table}[htbp]
\caption{Detection fraction  (ratio of the number of simulations with an SMBHB system detected to total simulations) for each observing strategy and population scenario, using an $(\mathrm{S/N})_{\mathrm{CW}}$ threshold of 5.94. ``TT Cons." here indicates that the total telescope time is conserved when making changes to the observing plan, so that fewer observations of the noisier pulsars are taken. ``Max. Cadence" denotes that the observational cadence of the given pulsars is maximized by reducing all other pulsar observations and conserving telescope time. Note that we implement a minimum of one observation every three months for all pulsars in the array.}
\centering
\begin{tabular}{l l c}
\toprule
\multicolumn{1}{c}{Scenario} & \multicolumn{1}{c}{Observing Strategy} & \multicolumn{1}{c}{\shortstack{Detection \\ Fraction}} \\
\midrule
\multirow{12}{*}{Heavy} & Fiducial (4.5 yr) & $0.06$ \\
 & Fiducial (9.0 yr) & $0.27$ \\
 & $2 \times$ Precision & $0.28$ \\
 & $2 \times$ Precision (TT Cons.) & $0.27$ \\
 & $4 \times$ Cadence & $0.30$ \\
 & $4 \times$ Cadence (TT Cons.) & $0.29$ \\
 & Combined & $0.33$ \\
 & Combined (TT Cons.) & $0.23$ \\
 & Max. Cadence -- Top 10 & $0.29$ \\
 & Max. Cadence -- Top 20 & $0.29$ \\
 & Max. Cadence -- Top 30 & $0.29$ \\
 & Max. Cadence -- Top 40 & $0.30$ \\
\addlinespace
\hline
\multirow{12}{*}{Intermediate} & Fiducial (4.5 yr) & $0.02$ \\
 & Fiducial (9.0 yr) & $0.16$ \\
 & $2 \times$ Precision & $0.17$ \\
 & $2 \times$ Precision (TT Cons.) & $0.16$ \\
 & $4 \times$ Cadence & $0.18$ \\
 & $4 \times$ Cadence (TT Cons.) & $0.17$ \\
 & Combined & $0.21$ \\
 & Combined (TT Cons.) & $0.13$ \\
 & Max. Cadence -- Top 10 & $0.18$ \\
 & Max. Cadence -- Top 20 & $0.18$ \\
 & Max. Cadence -- Top 30 & $0.18$ \\
 & Max. Cadence -- Top 40 & $0.18$ \\
\addlinespace
\hline
\multirow{12}{*}{Light} & Fiducial (4.5 yr) & $0.04$ \\
 & Fiducial (9.0 yr) & $0.10$ \\
 & $2 \times$ Precision & $0.11$ \\
 & $2 \times$ Precision (TT Cons.) & $0.11$ \\
 & $4 \times$ Cadence & $0.12$ \\
 & $4 \times$ Cadence (TT Cons.) & $0.12$ \\
 & Combined & $0.12$ \\
 & Combined (TT Cons.) & $0.08$ \\
 & Max. Cadence -- Top 10 & $0.11$ \\
 & Max. Cadence -- Top 20 & $0.11$ \\
 & Max. Cadence -- Top 30 & $0.12$ \\
 & Max. Cadence -- Top 40 & $0.12$ \\
\bottomrule
\end{tabular}
\label{tab:detection_fraction}
\end{table}

\begin{figure}[htbp]
 \begin{subfigure}{0.92\columnwidth}
    \vspace{-10pt}
     \includegraphics[width=\columnwidth]{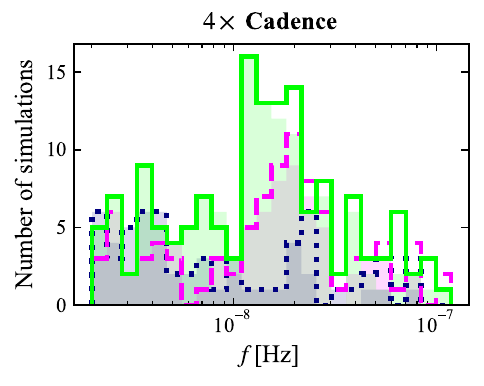}
     \label{fig:4cad_freqs}
 \end{subfigure}
 
 \vspace{-25pt}
 \begin{subfigure}{0.92\columnwidth}
    \vspace{-10pt}
     \includegraphics[width=\columnwidth]{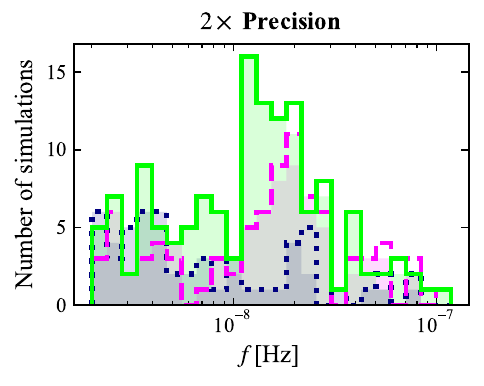}
     \label{fig:4prec_freqs}
 \end{subfigure}
 \vspace{-12pt}
 \begin{subfigure}{0.92\columnwidth}
    \vspace{-35pt}
     \includegraphics[width=\columnwidth]{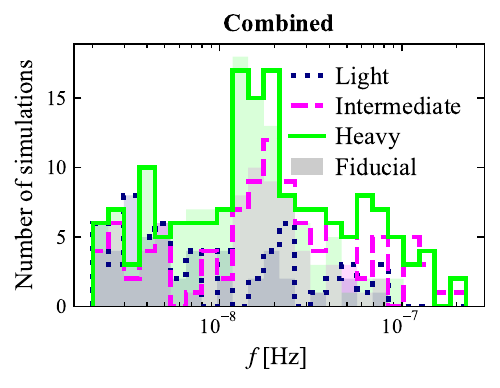}
     \label{fig:4cad_2prec_freqs}
 \end{subfigure}
  \caption{Distributions for the gravitational-wave frequency of the loudest detectable binary ($(\mathrm{S/N})_{\mathrm{CW}} \geq 5.94$) for a lighter, more numerous SMBHB population (dotted navy), an intermediate SMBHB population (dashed magenta), and a heavier, less numerous SMBHB population (solid lime). \textit{Top}: PTA with cadence increased by a factor of four for the ten pulsars contributing most to the $\rm (S/N)_{CW}$, \textit{middle}: PTA with precision increased by a factor of two for the ten pulsars contributing most to the $\rm (S/N)_{CW}$, \textit{bottom}: PTA with cadence increased by a factor of four and precision increased by a factor of two for the ten pulsars contributing most to the $\rm (S/N)_{CW}$. The shaded regions here show the results for the current MPTA observing plan extended to a 9.0-year data set.}
 \label{fig:synth_pta_freqs}
\end{figure}

We show the gravitational-wave frequencies of detectable sources for the higher cadence (top), higher precision (middle), and combined (bottom) simulated PTA data in Fig.~\ref{fig:synth_pta_freqs}.
As expected from the sensitivity curves in Fig.~\ref{fig:sensitivity_curve}, all synthesized PTAs have more sources at higher frequencies.
The shaded histograms show the fiducial observing strategy with the simulated MPTA 9.0-year PTA data.
The light (dotted navy) SMBHB population fluctuates but generally decreases with increasing frequency.
The intermediate (dashed magenta) case peaks at $15-20$ nHz for all scenarios, and the heavy (solid lime) SMBHB population peaks at $10-20$ nHz.  
All scenarios show an extended tail at higher frequencies compared with the fiducial observing strategy for each case.

Although Fig.~\ref{fig:sensitivity_curve} shows the most sensitive frequency is $\sim 6$ nHz, we find the median frequency is $5.2-17$ nHz for the higher precision strategy, $6.5-17$ nHz for higher cadence, and $6.5-20$ nHz for the combined case.
There is a clear trend of shifting to higher frequencies for the synthesized plans compared to the fiducial observing plan.
For each case, although the minimum frequency remains consistent across all PTAs, $\sim 2$ nHz, the sensitivity changes, and the distribution is spread to higher frequencies.
For the synthesized data, many of the highest $(\mathrm{S/N})_{\mathrm{CW}}$ sources were at higher frequencies compared with the fiducial observing plan forecasted over nine years.

\begin{figure}[htbp]
\includegraphics[width=\columnwidth]{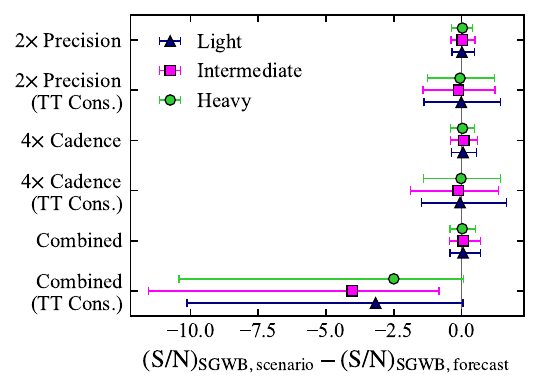}
  \caption{ The relative difference in the detection significance of the stochastic gravitational-wave background for various observing plans as compared with the current observing plan extended to nine years of data. The light population scenario is represented by navy triangles, the intermediate by magenta squares, and the heavy by lime circles. The error bars correspond to the 95\% credible intervals for each scenario and observing plan. ``TT Cons." here indicates that the total telescope time is conserved when making changes to the observing plan.}
\label{fig:SGWB_SN_scenarios}
\end{figure}

We show the stochastic background $(\mathrm{S/N})_{\mathrm{SGWB}}$ relative to the fiducial plan for all observing strategies in Fig.~\ref{fig:SGWB_SN_scenarios}.
We find that the $(\mathrm{S/N})_{\mathrm{SGWB}}$ is relatively unchanged across all strategies, except when conserving telescope time for the combined higher cadence and precision case.
For the cases where the telescope time is not conserved, $(\mathrm{S/N})_{\mathrm{SGWB}}$ is generally higher than the fiducial observing plan, as expected given that these scenarios are simply adding more observations to the existing fiducial plan.
When conserving telescope time with either higher cadence or higher precision, the ten most sensitive pulsars are observed for four times longer than in the fiducial case.
However, there is still sufficient time allocated to the remaining pulsars to establish a Hellings-Downs correlated signal, resulting in a marginally less optimal detection of the stochastic background as compared to the fiducial case.
For the combined case, we observe the most sensitive pulsars for sixteen times longer than the fiducial case, resulting in very few observations of the remaining, less-sensitive pulsars.
This leads to a marked reduction in the background detection with the telescope time conserved combined case, because many pulsars are required to optimize this detection.

\section{NANOGrav Candidates}
\label{sec:NG_comp}
We examine the distribution of detectable SMBHB systems as a function of frequency, chirp mass, and redshift.
This is relevant when determining the likelihood of candidate SMBHB systems from search pipelines.
\citet{agarwal_nanograv_2026} performed a targeted search for SMBHB systems in the NANOGrav 15-yr dataset and found two potential candidates.
The key properties of these candidates are shown in Table \ref{tab:ng_cands}.
\begin{table}[t]
    \caption{Properties of the two highest significance candidates from \citet{agarwal_nanograv_2026}.}
    \centering
    \begin{tabular}{c|c|c|c}
        Candidate & $z$ & $f$ [nHz] & $\log_{10}(\mathcal{M}/\rm{M}_{\odot})$ \\ \hline
        J0729+4008 & 0.074 & 14 & $9.38^{+0.11}_{-0.30}$ \\
        J1536+0441 &  0.379 & 21 & $9.67^{+0.11}_{-0.34}$\\
    \end{tabular}
    \label{tab:ng_cands}
\end{table}
As part of the analysis, they used population synthesis to generate populations consistent with the stochastic background detected in the NANOGrav 15-yr dataset (\citealt{agazie_nanograv_2023}) in a similar approach to ours.
Both candidates are consistent with their population synthesis predictions (see their Fig.~10).

\begin{figure*}[htbp]
     \begin{subfigure}{\textwidth}
     \includegraphics[width=\textwidth]{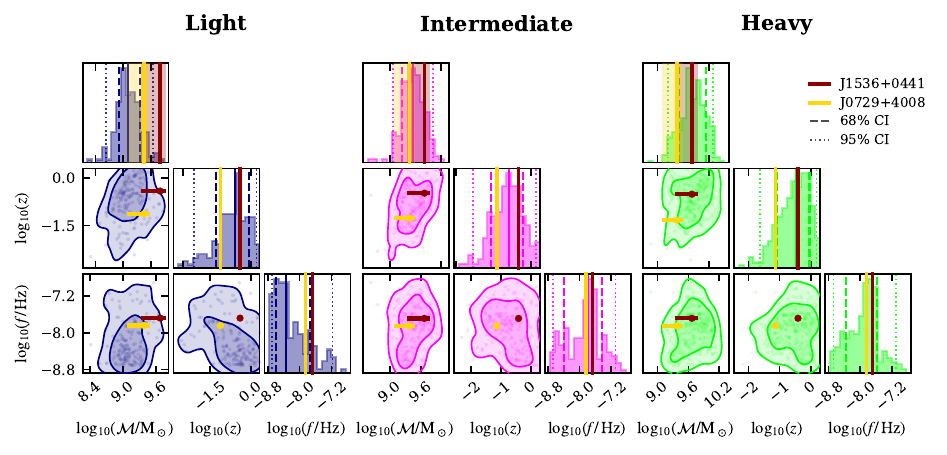}
 \end{subfigure}
    \caption{Distributions for the chirp mass, redshift, and observed gravitational-wave frequency for the light (\textit{left}), intermediate (\textit{middle}), and heavy (\textit{right}) scenarios. The dashed and dotted lines in the one-dimensional histograms, and the darker and lighter shaded regions in the joint parameter distributions show the 68\% and 95\% credible intervals from our analysis. The yellow and red denote the two candidates from \citet{agarwal_nanograv_2026}, where the shaded region denotes the listed 95\% credible interval for the parameter values, as does the length of the line in the joint parameter distributions. }
    \label{fig:corner_cands}
\end{figure*}

We analyze our simulated populations to determine if the candidate systems are plausible given our cosmological models.
Figure ~\ref{fig:corner_cands} shows the distributions for the chirp mass, redshift, and observed gravitational-wave frequency for the detectable binaries ($(\mathrm{S/N})_{\mathrm{CW}}\geq5.94$) in each of the light (left), intermediate (middle), and heavy (right) scenarios, respectively.
The two candidates from \citet{agarwal_nanograv_2026}, J0729+4008 and J1536+0441, are shown on the plots in yellow and red lines, respectively, where the length of the lines in the joint parameter distributions, and the shaded regions in the one-dimensional histograms correspond to the reported 95\% credible intervals.
The 68\% and 95\% credible intervals from our analysis are enclosed by the dashed and dotted lines, respectively, in the one-dimensional histograms.

We find that both candidates are consistent with our expectations for the first detectable binaries for all of our cosmological models, as the candidates' reported parameter ranges lie within the 95\% credible intervals for all parameters.
For the light scenario, the reported chirp mass of J1536+0441 is extreme compared to the simulated populations. However, when considering the lower limit from the reported 95\% credible interval, this falls within our population's chirp mass 95\% credible interval.
This indicates that although these candidates were consistent with noise \citep{agarwal_nanograv_2026}, these are the type of systems we expect to detect first with PTA data.

\section{Discussion} \label{sec:disc}

\subsection{Consequences for PTA Observing Campaigns}\label{subsec:campaigns}
For the current MPTA observing plan forecasted to a 9.0-year data set, detectable binaries are most likely to occur at frequencies of $\sim2-74\,$nHz. 
The low-frequency limit is set by the observing baseline, $f_{\mathrm{low}} \approx 1/T_{\mathrm{obs}}$, whereas the high-frequency limit is determined by the observing cadence, $f_{\mathrm{high}} \approx 2/\Delta t$, where $\Delta t$ is the cadence. 
The frequency of peak sensitivity depends on both cadence and timing precision; for example, the 4.5-year MPTA is most sensitive near $\sim10$\,nHz (Figure~\ref{fig:sensitivity_curve}), compared with $\approx6$\,nHz for the 15-year NANOGrav dataset, which considers frequencies of $\sim1-300\,$nHz (\citealt{arzoumanian_nanograv_2023}).

Previous efforts to inform PTAs of optimal cadences and observation strategies have focused on finding a balance between detecting the stochastic background and continuous-wave sources.
The background is stronger at lower frequencies, whereas at higher frequencies, there is less source confusion, allowing individual sources to be easier to detect. 
Given the emergence of evidence for this background, we propose that future observations may instead target higher frequencies to increase the likelihood of detecting individual SMBHB systems.

To be sensitive to these frequencies, pulsar timing observations can be shifted to higher cadence observations to maximize returns with telescope time. 
We show how important individual pulsars are in this analysis for detecting an individual SMBHB, and find that increasing the cadence of the most sensitive pulsars can raise the likelihood of detection.
We find that the median of the cumulative $(\mathrm{S/N})_{\mathrm{CW}}$ for each binary exceeds 50\% for the four most sensitive pulsars.
This indicates that increasing the cadence of the lowest-noise pulsars is the most promising method to increase the detection probability of SMBHB systems.

Shifting to higher cadence observations increases the probability of detecting a binary system to 12--30\% (with telescope time conserved, this is 12--29\%).
While increasing the precision of pulsar measurements increased the detection likelihood (11--28\%, or 11--27\% with telescope time conserved), increasing the cadence was more cost-effective for telescope time.
This is because timing precision scales as the square root of the duration, whereas for cadence this scales linearly.
Based on this work, we recommend that the MPTA increase the cadence of the most sensitive pulsars even if this means reducing observations of the remaining pulsars to conserve observing time.

In our analysis, we identify two subpopulations that are most likely to be detected as SMBHB systems: i) the intrinsically loud sources, that have high mass and are close by, with a high strain amplitude value, and ii) lower-mass and farther-distance sources, that have other parameters that influence the detector response dramatically, such as lower frequency and sky location proximal to the most sensitive sky location from the array.
Although further study is required to analyze the origins of these subpopulations in depth, this indicates that the first resolved sources may be less massive, but located in a more sensitive sky location and with a frequency that the PTA is more sensitive to, for example.

\subsection{Electromagnetic Counterparts} \label{subsec:EM_count}

High-frequency electromagnetically active continuous-wave sources can be monitored by current optical time-domain surveys such as the All-Sky Automated Survey for Supernovae (ASAS-SN; \citealt{shappee_man_2014, kochanek_all-sky_2017}), the Zwicky Transient Facility survey (ZTF; \citealt{bellm_zwicky_2018}), and the Asteroid Terrestrial-impact Last Alert System survey (ATLAS; \citealt{tonry_atlas_2018}).
These surveys observe half (ZTF) or all (ASAS-SN, ATLAS) of the sky, taking high-cadence observations (average cadence of $\sim2-3$ days, $\sim1$ day, and $\sim1-2$ days for ZTF, ASAS-SN, and ATLAS, respectively) over several years.
The long baselines and high cadences of these surveys provide an opportunity to observe periodic signatures on the timescales of months to years.
As most surveys have durations $\lesssim 10$ years, many of the longer-period ($\gtrsim$ a few years) candidates may be consistent with active galactic nuclei red noise processes as new data is added.
Irregular survey cadence may also lead to false positives in the SMBHB sample that decrease in statistical significance of the periodicity as new data is added.
This can lead to misidentification of single supermassive black holes as SMBHB systems due to accretion disk dynamics leading to quasi-periodicity (\citealt{vaughan_false_2016, liu_supermassive_2019, zhu_toward_2020, el-badry_active_2025, molina_search_2025}).

We use a threshold of observing $\gtrsim 5$ cycles of observations to distinguish true periodicity from red noise, as found in \citet{vaughan_false_2016}.
To verify such a candidate in a ten-year dataset (either using archival data or planned follow-up), one must focus on signals with frequencies of $\gtrsim\unit[32]{nHz}$, or equivalently optical variability on timescales of $\lesssim 2$ years.
While $\approx 85\%$ of the resolvable SMBHB systems in Fig.~\ref{fig:loudest_binary_threshold} fall below this threshold, we suggest that the 15\% of resolvable binaries with $f\gtrsim\unit[32]{nHz}$ are more valuable than lower-frequency detections because they are more practical to search for an electromagnetic counterpart.

There exist only a small number of SMBHB candidates from electromagnetic surveys of quasars that have periodicity on timescales $\lesssim 2$ years.
\citet{dorazio_observational_2023} provides a list of SMBHB candidates from various surveys, and out of the $\approx 300$ candidates (\citealt{graham_systematic_2015, charisi_population_2016_fixed, liu_supermassive_2019, chen_candidate_2020, chen_searching_2024}), $\approx 60$ have periods $\lesssim 2$ years.
The periodicity analysis is conducted on $\sim 4.3\times 10^5$ quasars, showing that $\approx 0.01\%$ of quasars may host SMBHB systems with periods $\lesssim 2$ years.
Although the short-period SMBHB candidate population constitutes a small fraction of the total SMBHB candidate population, the $\approx 60/300 = 20\%$ fraction of electromagnetic SMBHB candidates are the easiest to verify true periodicity over red noise compared to longer-period candidates.

\subsubsection{How distant can surveys observe SMBHB systems?}

We investigate the signature of SMBHB systems in electromagnetic data by assuming that our first candidate will actively accrete as a quasar.
We assume a typical quasar is accreting at an Eddington fraction of $\lambda = 0.1$, and that an SMBHB system may appear periodic in electromagnetic data through orbital Doppler boosting \citep[e.g.,][]{dorazio_observational_2023}.
Doppler boosting leads to periodicity in an SMBHB light curve due to the luminosity of a source changing as the line-of-sight velocity changes along the SMBHB orbit.
As a case study, we use the median detectable chirp mass for the intermediate case, $\sim 3 \times 10^9\;\rm{M}_{\odot}$, with a mass ratio $q \approx 0.5$, and an orbital period $P = 2$ years from the requirement of $5$ cycles to observe periodicity.
We use Equations 10--13 in \citet{dorazio_observational_2023} to calculate the change in magnitude, $\Delta m_{\rm{Doppler}}$, of the SMBHB due to the Doppler boosting over the binary period.

For a candidate to show periodicity in a light curve, the periodic amplitude must be much larger than the uncertainty from an observation in a given filter, $5\sigma_{m}$. 
As such, for this estimate, we require $\Delta m_{\rm{Doppler}} \geq 5\sigma_{m}$.
The uncertainty in the magnitude for a survey is dependent on the magnitude of the object that is being observed: near the limiting magnitude for ZTF, $m_{\mathrm{lim}} = 20.8$ for $g$-band, the uncertainty is $\sigma_{20.8} \approx 0.15$ \citep{bellm_zwicky_2018, RodriguezCastilloEtAl2020}, whereas at $m = 14$, this becomes $\sigma_{14} \approx 0.01$ \citep{masci_zwicky_2018}.
Between the limiting magnitude and the brighter magnitude for the associated uncertainty, we use a log-in-uncertainty-in-magnitude to linear-in-magnitude interpolation as a simple approximation.
A similar calculation can be made for ASAS-SN; however, ZTF has a fainter limiting magnitude than ASAS-SN \citep{inno_gaia-asas-sn_2021} using the same $g$-band, so we use ZTF to obtain a higher maximum observable distance.
For ATLAS, we assume at the limiting magnitude $m_{\mathrm{lim}} = 19.5$ in the $o$- or $c$-bands, $\sigma_{19.5} \approx 0.15$, and at $m = 14$, $\sigma_{14} \approx 0.015$ \citep{tonry_atlas_2018}.

Using these listed magnitudes and uncertainties in magnitudes, we can then solve for the point at which $\Delta m_{\rm{Doppler}} = 5\sigma_{m}$ from our interpolation of $\sigma_{m}$, which finds the faintest magnitude at which the Doppler boost causes clear periodic variations in the light curve.
We convert this apparent magnitude into a luminosity distance, finding that the maximum distance we could observe periodicity in these surveys is $\approx 3.2.0-6.3$ Gpc.
The upper end of this interval (corresponding to ZTF) is farther than the vast majority of the detectable binaries in Fig.~\ref{fig:loudest_binary_threshold}, while the lower end of the interval (corresponding to ATLAS) is farther than the majority of detectable binaries for all population scenarios.
This indicates that if these binaries are actively accreting, they are likely to be visible to these surveys, leading to the possibility of multi-messenger astronomy with the loudest binaries in PTA data and electromagnetic observations.

\subsection{Caveats}
In this work, we assume that SMBHB systems have circular orbits.
This simplifies the method, as introducing eccentricity leads to gravitational-wave emission at multiple frequencies from different harmonics for a given binary, $f = nf_{\mathrm{orb}}$, rather than solely $f = 2f_{\mathrm{orb}}$. 
This leads to a distribution of gravitational-wave energy from lower frequencies to higher frequencies.
Gravitational waves carry away angular momentum, rapidly circularizing the orbit \citep{peters_gravitational_1964}.
At lower frequencies, however, the binary is not necessarily circular.
Including eccentricity generally reduces $(\mathrm{S/N})_{\mathrm{CW}}$ for each binary, through the spreading of gravitational-wave power over frequencies.
There may be cases where spreading to higher frequencies may increase the $(\mathrm{S/N})_{\mathrm{CW}}$ for low-frequency binaries.

We test the effects of adding eccentricity on the $(\mathrm{S/N})_{\mathrm{CW}}$ and find that at low frequencies, $\lesssim 5$ nHz, there is an increase in $(\mathrm{S/N})_{\mathrm{CW}}$.
Above $\lesssim 5$ nHz, $(\mathrm{S/N})_{\mathrm{CW}}$ typically decreased with eccentricity.
The effect of eccentricity on continuous-wave signals was also studied in \citet{taylor_detecting_2016}, where they found that for $f > 5$ nHz, as $e\rightarrow 0.9$, $(\mathrm{S/N})_{\mathrm{CW}}$ approached 80\% of the non-eccentric case above 5 nHz.
For $e < 0.3$ ($< 0.5$), $(\mathrm{S/N})_{\mathrm{CW}}$ decreased by only $\sim 5\%$ ($\sim 10\%$).
Below 5 nHz they found that $(\mathrm{S/N})_{\mathrm{CW}}$ increased for all $e$ values.
This indicates that although including eccentricity will reduce the $(\mathrm{S/N})_{\mathrm{CW}}$, this impact is likely to be small in the context of this work.
Including eccentricity would also allow binaries with lower orbital frequency, below the PTA sensitivity, to radiate gravitational waves at higher frequencies into the MPTA band; however, the gravitational-wave contribution is dominated by the lower harmonics, such that only very loud binaries may be detected from this.
We save a systematic study of eccentricity for future analyses.

We also assume that the binaries simulated have no frequency evolution over the time span of observations.
Our source-frame frequency probability distribution, $p(f_s) \propto f_s^{-11/3}$, implicitly assumes frequency evolution in itself, as binaries are more slowly evolving at lower frequencies, and hence there will be more binaries in this regime.
When we inject the binary signals, we do not, however, evolve $f$ over observation time.
The gravitational-wave frequency evolves at a rate
\begin{equation}
    \dv{f}{t} = \frac{96}{5c^5} \pi^{8/3} (G\mathcal{M})^{5/3} f^{11/3}.
\end{equation}
Typically, this will be small, and we can estimate the shift in frequency (\citealt{sesana_measuring_2010})
\begin{equation}
    \begin{split}
        \Delta f \approx\;& \dv{f}{t} T_{\mathrm{obs}} \\ \approx\;& 0.05 \left(\frac{\mathcal{M}}{10^{8.5} \mathrm{M}_{\odot}} \right)^{5/3} \left(\frac{f}{50 \mathrm{nHz}} \right)^{11/3} \left(\frac{T_{\mathrm{obs}}}{10 \; \mathrm{yr}} \right) \; \mathrm{nHz}.
    \end{split}
\end{equation}
The combination of high frequency and high mass is rare, such that $\gtrsim 90\%$ of our loudest sources for all scenarios have a frequency shift of $< 1$ nHz.
While implementing frequency evolution should be explored, the results we have obtained will not be greatly affected by this.

\section{Conclusion} \label{sec:conc}
In this work, we examined what the characteristics of the first resolved supermassive black hole binary sources are likely to be. 
We find that these sources are expected to be massive ($\mathcal{M} \gtrsim 10^9\;\mathrm{M}_{\odot}$), relatively close ($z < 2$), and have a gravitational-wave frequency $2 \lesssim f \lesssim 75$ nHz.
The likelihood of detecting such systems in a simulated MeerKAT PTA 9.0-year data set with the 4.5-year observing strategy is $10-27\%$.
Increasing the cadence of the ten most sensitive pulsars in the array by a factor of four changes this likelihood to $12-33\%$, while increasing the precision of observations on these pulsars by a factor of two results in a detection likelihood of $11-28\%$.
Combining both of these strategies results in a detection probability of $12-33\%$.

Although these changes do not dramatically affect the detection probability, they do change which systems are likely to be detected, as these changes increase the sensitivity of the PTA at higher frequencies.
This means that higher gravitational-wave frequencies, or equivalently higher orbital frequencies, are more likely to be detected with these changes.
If SMBHB systems are electromagnetically active due to binary motion, higher-orbital-frequency systems will be easier to detect in current surveys because they will undergo more cycles in a given time.
Binaries with frequencies $\gtrsim 32$ nHz that are actively accreting as quasars are likely to be seen with current high-cadence transient surveys such as ZTF, ASAS-SN and ATLAS.

We recommend that pulsar timing arrays conduct higher-cadence observations of the most sensitive pulsars at the expense of decreasing observations on the less-sensitive pulsars in the array to both: a) increase the detection likelihood of continuous waves, and b) target the higher-frequency binaries for multi-messenger astronomy.
This will have a minimal effect on the likelihood of detecting the stochastic background.

\begin{acknowledgments}
We thank Valentina di Marco, Ryan Shannon, Riccardo Truant, and Saurav Mishra for useful discussions. 
Parts of this research were supported by the Australian Research Council Centre of Excellence for Gravitational Wave Discovery (OzGrav), through project number CE230100016.
The authors acknowledge the use of Claude (Anthropic) to optimize the code used to obtain the results in this work.
\end{acknowledgments}

\software{ \texttt{AstroPy} \citep{collaboration_astropy_2018}, \texttt{ENTERPRISE} \citep{enterprise}, \texttt{enterprise\_extensions} \citep{enterprise_ext}, \texttt{hasasia} \citep{Hazboun2019Hasasia}, \texttt{libstempo} \citep{vallisneri_libstempo_2020}, \texttt{MATPLOTLIB} \citep{4160265},   \texttt{NumPy} \citep{harris_array_2020}, \texttt{SciPy} \citep{virtanen_scipy_2020}. } 

\bibliographystyle{mnras}
{\scriptsize
  \bibliography{references, sample701}{}}
\end{document}